\documentstyle[12pt]{article}

\advance \topmargin by -\headheight
\advance \topmargin by -\headsep
     
\evensidemargin \oddsidemargin
\def\rf#1{(\ref{eq:#1})}
\def\lab#1{\label{eq:#1}}
\def\nonu{\nonumber}
\def\br{\begin{eqnarray}}
\def\er{\end{eqnarray}}
\def\be{\begin{equation}}
\def\ee{\end{equation}}
\def\eq{\!\!\!\! &=& \!\!\!\! }
\def\foot#1{\footnotemark\footnotetext{#1}}
\def\lb{\lbrack}
\def\rb{\rbrack}

\def\llb{\left\lbrack}
\def\rrb{\right\rbrack}

\def\lcurl{\left\{}
\def\rcurl{\right\}}
\def\({\left(}
\def\){\right)}
\def\v{\vert}                     
\def\bv{\bigm\vert}               
\def\bgv{\bigg\vert}              
\def\lskip{\vskip\baselineskip\vskip-\parskip\noindent}

\def\bc{\begin{center}}
\def\ec{\end{center}}

\newcommand{\sect}[1]{\setcounter{equation}{0}\section{#1}}

\relax
\newcommand\partder[2]{{{\partial {#1}}\over{\partial {#2}}}}

\newcommand\partderh[3]{{{\partial^{#3} {#1}}\over{{\partial {#2}}^{#3} }}}
\newcommand\partderm[3]{{{\partial^2 {#1}}\over{\partial {#2} \partial {#3} }}}

\newcommand\sbr[2]{\left\lbrack\,{#1}\, ,\,{#2}\,\right\rbrack} 
\newcommand\Sbr[2]{\Bigl\lbrack\,{#1}\, ,\,{#2}\,\Bigr\rbrack} 

\def\a{\alpha}
\def\b{\beta}

\def\d{\delta}
\def\D{\Delta}

\def\h{{1\over 2}}
\def\l{\lambda}

\def\m{\mu}

\def\o{\over}

\def\p{\phi}
\def\P{\Phi}
\def\pa{\partial}

\def\pr{\prime}

\def\s{\sigma}

\def\t{\tau}
\def\th{\theta}

\def\wti{\widetilde}
\newcommand\sumi[1]{\sum_{#1}^{\infty}}   

\newcommand\BDet[5]{\det_{{#1}}\left\Vert\begin{array}{cc}  
{#2} & {#3} \\ {#4} & {#5} \end{array} \right\Vert}   
\newcommand\Det[2]{\det_{{#1}} \left\Vert {#2} \right\Vert}

\def\cG{{\cal G}}
\def\cL{{\cal L}}
\def\cM{{\cal M}}
\def\cW{{\cal W}}
\def\phanta{\phantom{aaaaaaaaaaaaaaa}}
\font\numbers=cmss12
\font\upright=cmu10 scaled\magstep1
\def\stroke{\vrule height8pt width0.4pt depth-0.1pt}
\def\topfleck{\vrule height8pt width0.5pt depth-5.9pt}
\def\botfleck{\vrule height2pt width0.5pt depth0.1pt}
\def\Zmath{\vcenter{\hbox{\numbers\rlap{\rlap{Z}\kern 0.8pt\topfleck}\kern
2.2pt
                   \rlap Z\kern 6pt\botfleck\kern 1pt}}}
\def\Qmath{\vcenter{\hbox{\upright\rlap{\rlap{Q}\kern
                   3.8pt\stroke}\phantom{Q}}}}
\def\Nmath{\vcenter{\hbox{\upright\rlap{I}\kern 1.7pt N}}}
\def\Cmath{\vcenter{\hbox{\upright\rlap{\rlap{C}\kern
                   3.8pt\stroke}\phantom{C}}}}
\def\Rmath{\vcenter{\hbox{\upright\rlap{I}\kern 1.7pt R}}}
\def\IZ{\ifmmode\Zmath\else$\Zmath$\fi}
\def\IQ{\ifmmode\Qmath\else$\Qmath$\fi}
\def\IN{\ifmmode\Nmath\else$\Nmath$\fi}
\def\IC{\ifmmode\Cmath\else$\Cmath$\fi}
\def\IR{\ifmmode\Rmath\else$\Rmath$\fi}
\def\one{\hbox{{1}\kern-.25em\hbox{l}}}
\def\mark{\noindent{\bf Remark.}\quad}

\newtheorem{definition}{Definition}[section]
\newtheorem{proposition}{Proposition}[section]

\newtheorem{lemma}{Lemma}[section]
\newtheorem{corollary}{Corollary}[section]
\def\proof{\par{\it Proof}. \ignorespaces} \def\endproof{{$\Box$}\par}
 
\def\Win1{{\bf W_{1+\infty}}}           

\def\win1{{\bf w_{1+\infty}}}
\def\cKP{{\sf cKP}~}
\def\cKPrm{${\sf cKP}_{r,m}$~}

\newcommand\DB{{Darboux-B\"{a}cklund}~}
\def\Res{{\rm Res}}
\def\cKPro{${\sf cKP}_{r,1}$~}
\def\vp{{\varphi}}
\newcommand\BA{\psi_{BA} (t,\l)}   
\newcommand\BAc{\psi_{BA}^{\ast} (t,\l)}  
\newcommand\BAA[2]{\psi_{BA} \(#1, #2\)}   
\newcommand\BAAc[2]{\psi_{BA}^{\ast} \(#1, #2\)} 
\newcommand\X{{\widehat {\cal X}} \(\l, \mu\)}  

\newcommand{\ct}[1]{\cite{#1}}
\newcommand{\bi}[1]{\bibitem{#1}}
\newcommand\NPB[3]{{\sl Nucl. Phys.} {\bf B#1} (#2) #3}

\newcommand\CMP[3]{{\sl Commun. Math. Phys.} {\bf #1} (#2) #3}

\newcommand\PLA[3]{{\sl Phys. Lett.} {\bf #1A} (#2) #3}
\newcommand\PLB[3]{{\sl Phys. Lett.} {\bf #1B} (#2) #3}
\newcommand\JMP[3]{{\sl J. Math. Phys.} {\bf #1} (#2) #3}

\newcommand\SPTP[3]{{\sl Suppl. Prog. Theor. Phys.} {\bf #1} (#2) #3}

\newcommand\PR[3]{{\sl Phys. Reports} {\bf #1} (#2) #3}

\newcommand\FAaIA[3]{{\sl Functional Analysis and Its Application} {\bf #1}
(#2) #3}

\newcommand\LMP[3]{{\sl Letters in Math. Phys.} {\bf #1} (#2) #3}
\newcommand\IJMPA[3]{{\sl Int. J. Mod. Phys.} {\bf A#1} (#2) #3}
\newcommand\TMP[3]{{\sl Theor. Mat. Phys.} {\bf #1} (#2) #3}
\newcommand\JPA[3]{{\sl J. Physics} {\bf A#1} (#2) #3}

\newcommand\MPLA[3]{{\sl Mod. Phys. Lett.} {\bf A#1} (#2) #3}

\newcommand\PHSA[3]{{\sl Physica} {\bf A#1} (#2) #3}

\newcommand\JGP[3]{{\sl J. Geom. Phys.} {\bf #1} (#2) #3}

\begin{document}

\begin{titlepage}
\vspace*{-1.5cm}
\noindent
{\sl solv-int/9701017} \hfill{BGU-97/01/Jan-PH}\\
\phantom{bla}
\hfill{UICHEP-TH/97-1}\\
\begin{center}
{\large {\bf Method of Squared Eigenfunction Potentials in Integrable 
Hierarchies of KP Type}}
\end{center}
\vskip .15in
\begin{center}
{ H. Aratyn\footnotemark
\footnotetext{Work supported in part by the U.S. Department of Energy
under contract DE-FG02-84ER40173}}
\par \vskip .1in \noindent
Department of Physics \\
University of Illinois at Chicago\\
845 W. Taylor St., Chicago, IL 60607-7059, U.S.A.\\
{\em e-mail}: aratyn@uic.edu \\
\par \vskip .15in
{ E. Nissimov$^{2}$  and
S. Pacheva\foot{Supported in part by Bulgarian 
NSF grant {\em Ph-401}}}
\par \vskip .1in \noindent
Institute of Nuclear Research and Nuclear Energy \\
Boul. Tsarigradsko Chausee 72, BG-1784 $\;$Sofia, Bulgaria \\
{\em e-mail}: nissimov@inrne.acad.bg, svetlana@inrne.acad.bg \\
\begin{center} and \end{center}
Department of Physics, Ben-Gurion University of the Negev \\
Box 653, IL-84105 $\;$Beer Sheva, Israel \\
{\em e-mail}: emil@bgumail.bgu.ac.il, svetlana@bgumail.bgu.ac.il
\par \vskip .1in
\end{center}

\begin{abstract}
The method of squared eigenfunction potentials (SEP) is developed systematically
to describe and gain new information about Kadomtsev-Petviashvili (KP)
hierarchy and its reductions.
Interrelation to the $\t$-function method is discussed in detail.
The principal result, which forms the basis of our SEP method, is the
proof that any eigenfunction of the general KP hierarchy can be represented as
a spectral integral over the Baker-Akhiezer (BA) wave function with a spectral 
density expressed in terms of SEP. In fact, the spectral representations of the
(adjoint) BA functions can, in turn, be considered as defining equations for
the KP hierarchy. The SEP method is subsequently
used to show how the reduction of the full KP hierarchy
to the constrained KP ({\cKPrm}) hierarchies  can be given entirely in terms 
of linear constraint equations on the pertinent $\t$-functions.
The concept of SEP turns out to be crucial in providing a
description of \cKPrm hierarchies in the language of universal
Sato Grassmannian and finding the non-isospectral Virasoro symmetry generators
acting on the underlying $\t$-functions. The SEP method is used to 
write down {\em generalized binary} \DB transformations for 
constrained KP hierarchies whose orbits are shown to correspond to a new Toda
model on a {\em square} lattice. As a result, we obtain a series of new 
determinant solutions for the $\t$-functions generalizing the known Wronskian 
(multi-soliton) solutions. Finally, applications to random matrix
models in condensed matter physics are briefly discussed.
\end{abstract}

\end{titlepage}

\sect{Introduction}
\label{section:intro}

The primary object of this paper is the Kadomtsev-Petviashvili (KP) integrable 
hierarchy (for comprehensive reviews, see {\sl e.g.} \ct{Zakh,ldickey})
and its nontrivial reductions generalizing the familiar $r$-reduction
to the $SL(r)$ Korteweg-de Vries (KdV) hierarchy. 
The KP hierarchy is an infinite-dimensional system which admits different 
alternative formulations and exhibits many types of symmetries.
Here we are interested in a formulation based on the notion of
{\em squared eigenfunction potential}
and the spectral representations of the underlying eigenfunctions 
it gives rise to.
Because of its connection to vertex operators, many aspects of this theory
are algebraic in nature.
This allows us to discuss various symmetries of the hierarchy and applications 
to a large class of soliton systems obtained from it via symmetry 
reduction in a systematic manner.

The KP hierarchy arises as a set of compatibility conditions for the
linear spectral problem involving the pseudo-differential Lax operator
$\cL$ and the Baker-Akhiezer (BA) wave function $\BA$.
In recent years, the study of integrable systems of KP type
has undergone rapid growth due to the applications of the
tau-function technique invented by the Kyoto school \ct{KP,Miwa-Jimbo,hirota}.
The underlying principle of this method is to represent the relevant
soliton potentials and Hamiltonian densities in terms of isospectral flows
(with evolution parameters ~$(t) \equiv \(t_1 \equiv x ,t_2 ,\ldots \)$) of 
one single function $\t (t)$ in such a way that 
$\pa^2 \ln \t (t)/\pa t_1 \pa t_n $ becomes equal to the coefficient in 
front of $D^{-1}$ in the
pseudo-differential operator expansion of $\cL^n$.

In terms of the $\tau$-function, viewed as a function
of the infinitely many KP ``time''-variables $\(t_1 \equiv x ,t_2 ,\ldots \)$, 
the whole KP hierarchy is contained in Hirota's fundamental
bilinear identity \ct{hirota} instead of the infinite system of 
non-linear partial differential 
equations derived from the Sato-Wilson Lax operator approach.
The $\tau$-function approach bridges the way to several physical applications
in view of its direct connection to physical objects, such as correlation
and partition functions.
Moreover, it allows a coherent treatment of multi-soliton solutions.
These solutions of the nonlinear differential equations are generated
by the action of the Miwa-Jimbo vertex operator $\X$ \ct{Miwa-Jimbo}
(cf. eq.\rf{defx} below) on the $\t$-function. 
This vertex operator generates an infinitesimal B\"{a}cklund transformation 
of the KP hierarchy.
The family of all vertex operators constitutes a Lie algebra
isomorphic to $GL ({\infty})$. The transformation
$\tau(t) \to (\exp(a\X)) \tau(t)$ sends a solution of
the KP hierarchy into another solution. In this way, the action of
the infinite-dimensional Lie algebra  $GL ({\infty})$ on the solution space
of the KP equation is made explicit via the B\"{a}cklund transformation of
``adding one soliton".

This powerful formal machinery embeds many other concrete and
useful structures relevant for physical models. 
Recently a special class of solutions encountered in the matrix models
of discrete two-dimensional gravity was realized via the imposition of the 
Virasoro type of constraints on the underlying $\tau$-function
\ct{vira-constr,Morozov-Usp} (see also \ct{noak}).

A remarkable feature of the KP hierarchy is the existence of the
so called additional non-isospectral symmetries which, within the Lax
operator formalism, are generated by Orlov-Schulman pseudo-differential
operators \ct{Orlovetal}. The latter are
defined as purely pseudo-differential parts of products of powers of the Lax
operator $\cL$ and its ``conjugate'' $\cM$-operator 
(cf. eqs.\rf{add-symm-L},\rf{L-M} below) and their respective flows form
the infinite-dimensional Lie algebra $\Win1$\foot{$\Win1$ algebra was 
originally introduced in physics literature \ct{W-inf}
as a nontrivial ``large $N$'' limit of the associative, but {\em non}-Lie, 
conformal $\,{\bf W_N}$ algebra \ct{Zam}. It turns out to be isomorphic to
the (centrally extended) algebra of differential operators on the circle
\ct{dop}, {\sl i.e.}, the Lie algebra generated by $z^k (\pa/\pa z)^n$
for $k \in \IZ\,,\, n \geq 0$.
Let us also recall that the ``semiclassical'' limit (contraction)
${\bf w_{1+\infty}}$ of $\Win1$ is the algebra of area-preserving 
diffeomorphisms on the cylinder \ct{W-inf}.}.
In an important recent development Adler, Shiota, and van Moerbeke 
\ct{moerbeke,ASvM} (see also \ct{Dickey-addsym,cortona}) obtained a formula
for the KP hierarchy which 
relates the action of the vertex operator $\X$ on the $\tau$-function
to Orlov-Schulman non-isospectral additional symmetry flows on the BA wave 
function.
The coefficients in the spectral expansion of $\X$ act as vector fields on the 
space of $\t$-functions generating $\Win1$ algebra as well. Hence, 
the above result relates the $\Win1$ algebra acting on $\t (t)$ to the 
centerless $\Win1$ 
algebra of non-isospectral symmetry flows acting on the BA function $\BA$.

There exists an alternative to the $\t$-function method
characterization of the KP hierarchy evolution equations
in terms of (adjoint) eigenfunctions, {\sl i.e.}, functions whose KP multi-time
flows are governed by an infinite set of purely differential operators
$\lcurl B_k\rcurl_{k=1}^{\infty}$ (cf. Def.\ref{definition:eigen-def} below). 
The latter, by virtue of compatibility of the multi-time flows, have to satisfy 
the so called ``zero-curvature'' Zakharov-Shabat equations (cf. eq.\rf{ZS}
below). One can then show \ct{Tak} that all $B_k$ are obtained as purely
differential projections of $k$-th powers of a 
single pseudo-differential operator $\cL$, thus leading to the standard Lax
formulation of KP hierarchy. 

Overcoming the formal obstacle of having to define a function
via an inverse derivative $\pa_x^{-1}\,$ Oevel succeeded in \ct{oevela} to
associate a well-defined (up to a constant) function -- the squared
eigenfunction potential (SEP), to a pair of arbitrary eigenfunction and
adjoint eigenfunction such that the $x$-derivative of SEP coincides with
the product of the latter eigenfunctions.
Consequently, a systematic formalism emerged in \ct{oevela} for the study
of symmetries generated within the KP hierarchy via SEP \ct{orlov1}.
In a particular example, when both eigenfunctions defining the SEP are 
BA functions, the SEP becomes a generating function for the above mentioned 
additional non-isospectral symmetries of the KP hierarchy 
\ct{Orlovetal,moerbeke,ASvM,Dickey-addsym,cortona}.

In the SEP framework, the product of any pair of eigenfunction and adjoint
eigenfunction, being a $x$-derivative of SEP, can be viewed as a conserved
density within the hierarchy.
The transition to the important class of constrained KP hierarchies
${\sf cKP}_{r,m}$\foot{The \cKPrm integrable hierarchies appeared in 
different disguises from various parallel developments:
(a) symmetry reductions \ct{symm-constr,oevela,chengs} of the full {\sf KP} 
hierarchy; 
(b) abelianization, {\sl i.e.}, free-field realizations, in terms of finite 
number of fields, of both compatible first and second {\sf KP}
Hamiltonian structures \ct{multikp-m-2bose,Yu-no2rabn1-office};
(c) a method of extracting continuum integrable hierarchies from the 
generalized Toda-like lattice hierarchies \ct{BX93-94} underlying
(multi-)matrix models;
(d) purely algebraic approach in terms of the zero-curvature equations 
for the affine Kac-Moody algebras with non-standard gradations
\ct{lastyear}.},
which are Hamiltonian reductions of the general KP hierarchy and whose
Lax operators are given in eq.\rf{f-5} below,
can be effectuated by imposing equality between a linear combination of
$m$ ($m \geq 1$) conserved densities of the above mentioned type and the 
$r$-th ($r \geq 1$) fundamental Hamiltonian density of the KP hierarchy.
In such a case, the symmetry generated by SEP (called ``ghost'' flow) is
identified with the $r$-th isospectral flow of the original KP hierarchy.

The principal merit of our work is to reformulate the eigenfunction 
formalism of KP hierarchy 
in a new form called {\em squared eigenfunction potential (SEP) 
method}, 
namely, to employ SEP as a basic building block in defining the 
KP hierarchy. The main ingredient of the SEP method is the proof of existence
of spectral representation for any eigenfunction involving SEP
as an integration kernel (spectral density).
A link is then provided between the two alternative formulations of
the KP hierarchy: one based on the $\t$-function and another one based on the 
SEP method. 
Furthermore, we apply the SEP method to solve various issues in 
integrable models of KP type and their applications in physics, among them, 
deriving new determinant solutions
for the $\t$-function containing the familiar Wronskian (multi-soliton)
solutions as simple particular cases, as well as identifying them
as possible novel types of joint distribution functions
in random matrix models of condensed matter physics.

The plan of the paper is as follows.
After reviewing the background material in Section \ref{section:survey},
we first prove in Section \ref{section:spectral} that any eigenfunction of 
the general KP hierarchy can be represented as a spectral integral over the 
BA wave function with a spectral density expressed in terms of SEP.
When (at least one) of the eigenfunctions is a BA functions, we obtain a 
closed expression for the SEP.
When both of the eigenfunctions are BA functions, the resulting SEP's
are connected straightforwardly to the bilocal vertex operator $\X$ 
acting on the $\t$-function. This association leads to a 
simple alternative proof for the Adler, Shiota, and 
van Moerbeke result \ct{moerbeke,ASvM,Dickey-addsym,cortona}
mentioned above.

A further important observation in Section \ref{section:spectral} is that
the spectral representation equations for the (adjoint) BA functions themselves
can be considered as defining equations for the KP hierarchy.
In other words, our formalism of spectral representations of KP eigenfunctions 
can be viewed as an equivalent alternative characterization of the KP hierarchy
parallel to Hirota's bilinear identity or Fay's trisecant identity
\ct{moerbeke}.

Our results in the constrained \cKPrm hierarchy case are as follows.
In Section \ref{section:grassmann}, using the SEP framework we obtain
an equivalent description (parallel to the one within the Lax
pseudo-differential operator approach) of \cKPrm hierarchies entirely in terms
of $\t$-functions only. Namely, we first derive a linear equation for the
$\t$-function (eq.\rf{tau-t-n-X} below), involving the bilocal vertex operator
$\X$ and a set of spectral densities, which uniquely constrains the
$\t$-function to belong to the \cKPrm hierarchy.
Furthermore, we provide  in Section \ref{section:grassmann}
an alternative description of \cKPrm hierarchies in the language of 
universal Sato Grassmannian.

One of the advantages of the SEP approach lies in the fact that it allows for
a coherent treatment of the non-isospectral symmetries for
KP-type hierarchies.
We use this feature in Section \ref{section:virasoro} to demonstrate how
the combination of the familiar Orlov-Schulman non-isospectral
symmetry flows, operating in the full unconstrained KP hierarchy,
together with certain appropriately chosen additional SEP-generated ``ghost''
symmetry flows \ct{addsym,noak}
gives rise to the correct non-isospectral Virasoro symmetry generators
acting on the space of \cKPrm $\t$-functions\foot{The standard
Orlov-Schulman non-isospectral symmetry flows do {\em not} preserve the
constrained form \rf{f-5} of \cKPrm hierarchy.}.

The SEP method is applied further in Section \ref{section:binary} to
formulate {\em generalized multi-step binary} \DB (DB) transformation rules 
of (constrained) KP hierarchies 
(one-step binary DB transformations with SEP have been introduced previously 
in ref.\ct{oevelb}).
Based on these transformation rules and using the fundamental Fay identities, 
we derive a series of new determinant solutions for the $\t$-functions 
generalizing the known Wronskian (multi-soliton) solutions.
The binary DB orbits define a discrete symmetry structure for \cKPrm
hierarchies corresponding to a square lattice. We exhibit the equivalence of
these binary DB orbits with a generalized Toda model on a {\em square} lattice.

Our formalism offers applications to the study of random matrix
models in condensed matter physics, which we briefly discuss in Section 
\ref{section:random}, where we also present some discussion of the results
and an outlook.

\sect{Background on General and Constrained KP Hierarchies}
\label{section:survey}
The calculus of the pseudo-differential operators (see e.g. \ct{KP,ldickey}) 
is one of the principal approaches to describe the KP hierarchy of integrable 
nonlinear evolution equations. In what follows the operator $D$ is such that
$ \sbr{D}{f} = \pa f = \pa f /\pa x$ and the
generalized Leibniz rule holds:
\be
D^n f  = \sumi{j=0} {n \choose j} (\pa^j f) D^ {n-j}  \quad , \quad
n \in \IZ
\lab{gleib}
\ee
In order to avoid confusion we shall employ the following notations:
for any (pseudo-)\-differential operator $A$ and a function $f$, the symbol
$\, A(f)\,$ will indicate application (action) of $A$ on $f$, whereas the
symbol $Af$ will denote just operator product of $A$ with the zero-order
(multiplication) operator $f$.

In this approach the main object is the pseudo-differential {\em Lax} 
operator:
\be
L = D^r + \sum_{j=0}^{r-2} v_j D^j + \sum_{i=1}^{\infty} u_i D^{-i}
\lab{lax-op}
\ee
The Lax equations of motion:
\be
\partder{L}{t_n} = \lb L^{n/r}_{+} \, , \, L \rb \quad , \; \;
n = 1, 2,
\ldots
\lab{lax-eq}
\ee
describe isospectral deformations of $L$. In \rf{lax-eq} and in what follows
the subscripts $(\pm )$ of any pseudo-differential
operator $A = \sum_j a_j D^j$ denote its purely differential part
($A_{+} = \sum_{j\geq 0} a_j D^j$) or its purely pseudo-differential part
($A_{-} = \sum_{j \geq 1} a_{-j} D^{-j}$), respectively. Further,
$(t) \equiv (t_1 \equiv x, t_2 ,\ldots )$ collectively  denotes the infinite
set of evolution parameters (KP ``multi-time'') from \rf{lax-eq}.

The Lax operator \rf{lax-op} can be represented equivalently in terms of the
so called {\em dressing} operator $W$ :
\be
W= 1 + \sum_{n=1}^{\infty} w_n D^{-n} \qquad ; \qquad
L = W D^r\,W^{-1}
\lab{dress-1}
\ee
whereupon the Lax eqs.\rf{lax-eq} become equivalent to the so called 
Wilson-Sato equations:
\be
\frac{\pa W}{\pa t_n} = L^{n/r}_{+} W - W D^n
= - L^{n/r}_{-} W
\lab{sato-a}
\ee
Further important object is the Baker-Akhiezer (BA) ``wave'' function 
defined via:
\be
\psi_{BA} (t,\l ) = W\bigl( e^{\xi (t,\l )}\bigr) = w(t,\l )e^{\xi  (t,\l )}
\quad ; \quad
w(t,\l ) = 1 + \sum_{i=1}^{\infty} w_i(t)\l^{-i} \ ,
\lab{BA}
\ee
where
\be
\xi(t,\lambda) \equiv  \sum_{n=1}^\infty t_n\lambda^n \qquad; \quad
t_1 = x
\lab{xidef}
\ee
Accordingly, there is also an adjoint BA function:
\be
\psi_{BA}^{*} (t,\l ) = W^{*-1}\bigl( e^{-\xi (t,\l )}\bigr) =
w^{*}(t,\lambda) e^{-\xi (t,\l )}     \quad ;\quad
w^{*}(t,\lambda) = 1 + \sum_{i=1}^{\infty} w_i^{*}(t)\lambda^{-i}
\lab{BA-adjoint}
\ee
The (adjoint) BA functions obey the following linear system:
\br
L\bigl( \psi_{BA} (t,\l )\bigr) &= &\l^r \psi_{BA} (t,\l ) \quad , \quad
\partder{}{t_n} \BA = L^{n/r}_{+}\bigl(\psi_{BA} (t,\l )\bigr)
\lab{linsys} \\
L^{*}\bigl( \psi_{BA}^{*} (t,\l )\bigr) &= & \l^r\psi_{BA}^{*} (t,\l )
\quad , \quad
\partder{}{t_n} \BAc  =
- \( L^{\ast}\)^{n/r}_{+}\bigl(\psi_{BA}^{*}(t,\l )\bigr)      \nonu
\er
Note that eqs.\rf{lax-eq} for the KP hierarchy flows can be regarded as
compatibility conditions for the system \rf{linsys}.

There exists another equivalent and powerful approach to the KP
hierarchy based on one single function of all evolution parameters --
the so called tau-function $\tau(t)$ \ct{KP} . It
is an alternative to using the Lax operator and the calculus of the
pseudo-differential operators. The $\t$-function is related to the BA functions
\rf{BA}--\rf{linsys} via:
\br
\psi_{BA}(t,\l) \eq
\frac{\tau \bigl( t - [\l^{-1}]\bigr)}{\tau (t)} e^{\xi(t,\l)}
= e^{\xi(t,\l)} \sumi{n=0} \frac{p_n \( - [\pa]\)\tau (t)}{\tau (t)} \l^{-n}
\lab{psi-main} \\
\psi_{BA}^{*}(t,\l) \eq
\frac{\tau \bigl(t + [\l^{-1}]\bigr)}{\tau (t)} e^{-\xi(t,\l)}
= e^{-\xi(t,\l)} \sumi{n=0} \frac{p_n \( [\pa]\)\tau (t)}{\tau (t)}
\l^{-n}
\lab{psi-mainc}
\er
where:
\be
[\l^{-1}] \equiv \( \l^{-1}, \h \l^{-2}, {1\over 3} \l^{-3},\dots\)
\qquad ; \qquad
[\pa] \equiv \(\partder{}{t_1}, \h \partder{}{t_2}, {1\over 3} \partder{}{t_3},
\ldots\)
\lab{tau-short-hand}
\ee
and the Schur polynomials $p_n (t)$ are defined through:
\be
e^{\sum_{l \geq 1} \l^l t_l} =
\sumi{n=0} \l^n \, p_n (t_1, t_2, \ldots )
\lab{Schur}
\ee
Taking into account \rf{psi-main} and \rf{BA}, the expansion for the
dressing operator \rf{dress-1} becomes:
\be
W = \sum_{n=0}^{\infty} \frac{p_n \( - [\pa]\)\tau (t)}{\tau (t)} D^{-n}
\quad , \quad i.e. \;\; w_1 (t) = \Res W = - \pa_x \ln \t (t)
\lab{W-main}
\ee

The (adjoint) BA functions enter the fundamental Hirota bilinear identity:
\be
\int  d\l \,\psi_{BA}(t,\l) \psi_{BA}^{*}(t',\l) = 0
\lab{bilide}
\ee
which generates the entire KP hierarchy. 
Here and in what follows integrals over spectral parameters are 
understood as: 
$\int d\l \equiv \oint_{0} \frac{d\l}{2i\pi} ={\rm Res}_{\l = 0}$.

Let us also recall that the KP hierarchy possesses an infinite set of
commuting integrals of motion w.r.t. the compatible first and second
fundamental Poisson-bracket structures whose densities
$h_{l-1} = {1\o l} {\rm Res} L^{l/r}$ are expressed in terms of the
$\t$-function \rf{psi-main} as:
\be
\pa_x \partder{}{t_l}\ln \t (t) = {\rm Res} L^{l/r}
\lab{tau-L}
\ee

Below we shall be particularly interested in reductions of the full
(unconstrained) KP hierarchy \rf{lax-op}. In this respect, it turns out that
a crucial r\^{o}le is played by the notions of {\em (adjoint) eigenfunctions}
of KP hierarchy.
\begin{definition}
The function $\Phi$ ($ \Psi$) is called \underbar{(adjoint) eigenfunction} of
the Lax operator $L$ satisfying Sato's flow equation \rf{lax-eq} if its
flows are given by the expressions:
\be
\partder{\Phi}{t_k} = L^{k/r}_{+}\bigl( \Phi\bigr) \qquad; \qquad
\partder{\Psi}{t_k} = - \( L^{*} \)^{k/r}_{+}\bigl( \Psi\bigr)
\lab{eigenlax}
\ee
for the infinitely many times $t_k$.
\label{definition:eigen-def}
\end{definition}
Of course, according to \rf{linsys} the (adjoint) BA functions are 
particular examples of (adjoint) eigenfunctions which, however, satisfy
in addition also the corresponding spectral equations.

In what follows a very important r\^{o}le will be played by the notion of
the so called {\em squared eigenfunction potential} (SEP).
As shown by Oevel
\ct{oevela}, for an arbitrary pair of (adjoint) eigenfunctions
$ \P (t), \Psi (t) $ there exists the function $S \( \P (t), \Psi (t) \)$,
called SEP, which possesses the following characteristics:
\be
{\pa \o \pa t_n} S \( \P (t), \Psi (t) \) = 
{\rm Res} \( D^{-1} \Psi (L^{n/r})_{+} \P D^{-1} \)
\lab{potentialflo}
\ee
The argument in \ct{oevela}, proving the existence of $S\(\P (t),\Psi (t)\)$,
was built on compatibility between isospectral
flows as defined in eq. \rf{potentialflo} and \rf{eigenlax}.

In particular, for $n=1$ eq.\rf{potentialflo} implies that the space 
derivative (recall $\pa_x \equiv \pa/\pa t_1$) of $S \( \P (t), \Psi (t) \)$
is equal to the product of the underlying eigenfunctions:
\be
\pa_x S \( \P (t), \Psi (t) \) =   \P (t)\,  \Psi (t)
\lab{potentialx}
\ee
\mark
Eq.\rf{potentialflo} determines $S\(\P (t),\Psi (t)\)$ up to a shift by a
trivial constant.

{}From eqs.\rf{potentialflo}--\rf{potentialx} one sees that $\P (t)\, \Psi (t) $
is a conserved density of the KP hierarchy.
This fact has a special significance for the reduction of the general
KP hierarchy to the constrained \cKPrm models (see below).
\begin{definition}
The constrained KP hierarchy (denoted as ${\sf cKP}_{r,m}$) is given by a
Lax operator of the following special form:
\be
L \equiv L_{r,m}  = D^r+ \sum_{l=0}^{r-2} u_l D^l +
\sum_{a=1}^m \Phi_a D^{-1} \Psi_a
\lab{f-5}
\ee
\label{definition:cKPrm}
\end{definition}
One can easily check that the functions $\Phi_a ,\Psi_a$, entering the
purely pseudo-differential part of $L_{r,m}$ \rf{f-5}, are (adjoint)
eigenfunctions of $L_{r,m}$ according to Def.\ref{definition:eigen-def}.

The purely pseudo-differential part of arbitrary power of the \cKPrm Lax
operator \rf{f-5} has the following explicit form \ct{EOR95}:
\be
\(  L^k \)_{-} = \sum_{a=1}^m \sum_{j=0}^{k-1} L^{k-j-1} (\Phi_a) D^{-1}
\( L^{\ast}\)^{j} (\Psi_a  )
\lab{lkminus}
\ee
Formula \rf{lkminus} can easily be derived from the simple technical
identity involving a product of two pseudo-differential operators
of the form ~$ f_i D^{-1} g_i\;, \, i=1,2$ :
\be
f_1 D^{-1} g_1  f_2 D^{-1} g_2
 = f_1 S(f_2 ,g_1 ) D^{-1} g_2 - f_1 D^{-1} S(f_2 ,g_1 ) g_2
\lab{x1x2}
\ee
where $f_i ,g_i$ are pairs of (adjoint) eigenfunctions of some KP Lax
operator, with $S(\cdot ,\cdot )$ being the corresponding SEP.

Note, that in agreement with eq.\rf{x1x2} the expression $L(\P_a )$
in \rf{lkminus} with $L$ as in \rf{f-5} explicitly reads:
$L(\P_a ) = L_{+} (\P_a ) + \sum_{b=1}^m \P_b S \( \P_a, \Psi_b\)$.
This definition extends naturally to higher powers of $L$ acting on $\P_a$
as well as higher powers of $L^{\ast}$ acting on $\Psi_a$.
Moreover, one can easily show that $L^l \(\P_a\)$ and
${L^\ast}^k \(\Psi_a\)$ are (adjoint) eigenfunctions of $L$ \rf{f-5} as well.

For three pseudo-differential operators
$ X_i \equiv f_i D^{-1} g_i\;, \, i=1,2,3$ the associativity law 
$\( X_1 X_2 \) X_3=  X_1 \(X_2  X_3\)$ implies the following technical Lemma:
\begin{lemma}
The squared eigenfunction potential $S (\cdot, \cdot)$ satisfies:
\be
S (f,g) S (h,k) = S \( h, k S (f,g)\) + S \( f S (h,k) , g\)
\lab{associat}
\ee
for (adjoint) eigenfunctions $f,g,h,k$.
\label{lemma:associati}
\end{lemma}
\sect{Spectral Representation of KP Eigenfunctions}
\label{section:spectral}
Consider the bilocal vertex operator \ct{Miwa-Jimbo} :
\br
\X &\equiv&
{1 \o \l} : e^{{\hat \th}(\l )}:\, : e^{-{\hat \th}(\m )}: =
{1 \o \l} e^{\xi \( t+[\l^{-1}], \m \)- \xi(t,\l)}\, e^{\sumi{1} {1 \o l}
\( \l^{-l} - \m^{-l} \) \partder{}{t_l} } \nonu\\
&=& - {1 \o \m} e^{\xi \( t, \m \)- \xi(t-[\m^{-1}],\l)}\,
e^{\sumi{1} {1 \o l} \( \l^{-l} - \m^{-l} \) \partder{}{t_l} } + \d (\l, \m)
\lab{defx}
\er
where:
\be
{\hat \th}(\l ) \equiv - \sum_{l=1}^{\infty} \l^l {t_l} +
\sum_{l=1}^{\infty} {1\o l} \l^{-l} \partder{}{t_l}
\lab{theta-def}
\ee
$\xi \( t, \l\)$ is as in \rf{xidef}, the columns
$: \ldots :$ indicate Wick normal ordering w.r.t. the creation/annihilation
``modes'' $t_l$ and $\partder{}{t_l}$, respectively, and the delta-function
is defined as:
\be
\d (\l, \m) = {1 \o \l} {1 \o 1- {\m \o \l }} +
{1 \o \m} {1 \o 1- {\l \o \m }}
\lab{defdelta}
\ee
An equivalent representation for $\X$, using Wick theorem, reads:
\be
\X = {1 \o {\l -\m}} : e^{{\hat \th}(\l ) - {\hat \th}(\m )}: =
{1 \o \l -\m}\, e^{\xi \( t, \m \)- \xi(t,\l)} \, e^{\sumi{1} {1 \o l}
\( \l^{-l} - \m^{-l} \) \partder{}{t_l} }
\quad {\rm for} \quad \v \m \v \leq \v \l \v
\lab{defx1}
\ee
The vertex operator $\X$ can be expanded as follows:
\be
\X = {1 \o {\l -\m}}  \sum_{l=0}^{\infty} \frac{(\m -\l )^l}{l!}
\sum_{s=-\infty}^{\infty} \l^{-s-l-1} {1\o{l+1}} {\widehat W}^{(l+1)}_s
\lab{X-expand}
\ee
where the operators ${\widehat W}^{(l)}_s$ span $\Win1$ algebra.

{}From the standard representation for the (adjoint) Baker-Akhiezer wave function
\rf{psi-main},\rf{psi-mainc} in terms of the $\t$-function we deduce the
identity:
\br
{ \X \t (t) \o \t (t)} &=& {1 \o \l} \, \BAc \, \BAA{t+[\l^{-1}]}{\m}
\lab{xidentitya} \\
&=& - {1 \o \m}\,  \BAA{t}{\m} \, \BAAc{t-[\m^{-1}]}{\l} + \d (\l, \m)
\lab{xidentityb}
\er
Now, recall the Fay identity \ct{moerbeke} for $\t$-functions:
\be
(s_0-s_1)(s_2-s_3)\tau(t+[s_0]+[s_1])\tau(t+[s_2]+[s_3])+{\rm
cyclic}(1,2,3)=0
\lab{fayid}
\ee
which, in fact, is equivalent to Hirota bilinear identity \rf{bilide}.
In what follows, we shall often make use of a particular form of \rf{fayid}
upon setting $s_0 = 0$, dividing by $s_1 s_2 s_3$ and shifting the KP times
$(t) \to \( t - [s_2] - [s_3]\)$ :
\br
\( s_2^{-1} - s_3^{-1}\) \t\( t+[s_1]-[s_2]-[s_3]\) \t\( t\) +
\( s_1^{-1} - s_2^{-1}\)\t \( t-[s_2]\) \t\( t+[s_1]-[s_3]\)  \nonu \\
+ \( s_3^{-1} - s_1^{-1}\)\t \( t-[s_3]\) \t\( t+[s_1]-[s_2]\) = 0
\lab{fayid-0}
\er
Especially, making identification $s_1= \m^{-1}$, $s_2= z^{-1}$
and $s_3= \l^{-1}$ in \rf{fayid-0} and using \rf{psi-main}--\rf{psi-mainc},
we arrive at the following useful Lemma:
\begin{lemma}
The truncated Fay identity \rf{fayid-0} is equivalent to the following
bilinear identity for (adjoint) BA functions:
\be
{1\o \l}{\widehat \D}_{z} \Bigl(\BA \psi_{BA}^{\ast}(t-[\l^{-1}],\m)\Bigr) =
- {1 \o z} \psi_{BA} (t,\l) \psi_{BA}^{\ast} (t-[z^{-1}],\m)
\lab{delpsi2}
\ee
where ${\widehat \D}_{z}$ is the shift-difference operator acting on
functions depending on the variables $t=(t_1,t_2,...)$ as follows:
\be
{\widehat \D}_{z} \equiv e^{\sumi{1} {1\o l}z^{-l}\pa/\pa t_l} - 1
\quad ,\quad
{\widehat \D}_{z} f(t) = f\( t-[z^{-1}]\) - f(t)
\lab{deltal}
\ee
\label{lemma:deltaonpsi}
\end{lemma}

The Fay identity \rf{fayid} is also known in its differential version:
\be
\pa_x \( {\tau \(t+[\l^{-1}] -[\m^{-1}]\) \o \tau (t)} \)
= \( \l - \m \) \( {\tau \(t+[\l^{-1}] -[\m^{-1}]\) \o \tau (t)}
- {\tau \(t+[\l^{-1}] \) \o \tau (t)}{\tau \(t-[\m^{-1}] \) \o \tau (t)}\)
\lab{dfayid}
\ee
Using \rf{defx1} and multiplying both sides of \rf{dfayid}
by $\exp\lcurl - \xi \( t, \l\) + \xi \( t,\m \)\rcurl$ we can rewrite it as:
\be
\pa_x \( {\X  \tau \(t\) \o \tau (t)} \)
= - \BAc \BAA{t}{\m}
\lab{fayx}
\ee
or, equivalently, using \rf{xidentitya} and \rf{xidentityb} :
\br
\pa_x \(- {1 \o \l} \, \BAc \, \BAA{t+[\l^{-1}]}{\m} \) &=&
\BAc \BAA{t}{\m} \nonu \\
\pa_x \( {1 \o \m}\,  \BAA{t}{\m} \, \BAAc{t-[\m^{-1}]}{\l}\)&=&
\BAc \BAA{t}{\m}
\lab{xidentityc}
\er

Let $\P, \Psi$ be a pair of an eigenfunction and an adjoint eigenfunction
of the general KP hierarchy. Our main statement in this section is:
\begin{proposition}
Any (adjoint) eigenfunction of the general KP hierarchy possesses a spectral
representation of the form:
\be
\P (t) = \int d \l \, \vp (\l)\, \BA \quad ; \quad
\Psi (t) = \int d \l \, \vp^{\ast} (\l)\, \BAc  \lab{spec}
\ee
with spectral densities given by:
\be
\vp (\l)  = { 1\o \l} \BAAc{t^{\pr}}{\l} \P \(t^{\pr}+ [\l^{-1}]\)
\quad ; \quad
\vp^{\ast} (\l)  = { 1\o \l} \BAA{t^{\pr}}{\l}
\Psi \(t^{\pr}- [\l^{-1}]\)
\lab{specdens}
\ee
where the multi-time $t^{\pr} = \( t_1^{\pr},t_2^{\pr},\ldots \)$
is taken at some arbitrary fixed value. In other words:
\br
\P (t) &=& \int d \l\, \BA { 1\o \l} \BAAc{t^{\pr}}{\l}
\P \(t^{\pr}+ [\l^{-1}]\)
\lab{spec1}\\
\Psi (t) &=& \int d \l \, \BAc { 1\o \l} \BAA{t^{\pr}}{\l}
\Psi \(t^{\pr}- [\l^{-1}]\)
\lab{spec2}
\er
are valid for arbitrary KP (adjoint) eigenfunctions $\P , \Psi$ and
for an arbitrary fixed multi-time $t^{\pr}$. Furthermore, the r.h.s. of
\rf{spec1} and \rf{spec2} do not depend on $t^{\pr}$.
\label{proposition:propspec}
\end{proposition}
We will proceed proving the above proposition in two steps.
First, let us assume that the (adjoint-)eigenfunctions indeed possess a
spectral representation of the form \rf{spec} with some spectral densities
$\vp^{(\ast )}(\l )$ . In such case the statement of the proposition is
contained in a much simpler Lemma:
\begin{lemma}
For (adjoint) eigenfunctions possessing the spectral representation 
\rf{spec} their
respective spectral densities are given by \rf{specdens}.
Consequently, in this case the equations \rf{spec1} and \rf{spec2}
are valid too.
\label{lemma:spec}
\end{lemma}
\begin{proof}
Using the spectral representation \rf{spec} for $\P \(t^{\pr}+ [\l^{-1}]\)$
and substituting it into the right hand side of \rf{spec1}, we get:
\be
\int d \l \int d \m  \, \vp (\m)\, \BA { 1\o \l} \BAAc{t^{\pr}}{\l}
\BAA{t^{\pr}+ [\l^{-1}]}{\m}
\lab{specpr1}
\ee
Recalling \rf{xidentityb} we can rewrite \rf{specpr1} as:
\br
&&\int d \l \int d \m \,  \vp (\m)\, \BA \({- 1\o \m} \BAA{t^{\pr}}{\m}
\BAAc{t^{\pr}- [\m^{-1}]}{\l} + \d \( \l , \m\)\) \nonu\\
&=& \int d\l\, \vp (\l)\, \BA = \P (t)
\lab{specpr2}
\er
where use was made of the fundamental Hirota bilinear identity \rf{bilide}.
The $t^{\pr}$-independence of the r.h.s. of \rf{spec1} and \rf{spec2} will be
demonstrated explicitly in the course of proof of
Prop.\ref{proposition:propspec} given below.
\end{proof}
We are now ready to take a final step of the proof of
Prop.\ref{proposition:propspec} and
extend the result of Lemma (\ref{lemma:spec}) to arbitrary
KP (adjoint-)eigenfunctions {\em without} assuming existence of a spectral
representation \rf{spec}.
To this end we need to recall the notion of SEP
\rf{potentialflo}--\rf{potentialx}.

Let $ S\(\P (t),\BAc \)$ be the SEP associated
with a pair of eigenfunctions $\P (t)$ and $\BAc$, {\sl i.e.}
$\pa_x S\( \P (t), \BAc \) = \P (t) \BAc$. Define now:
\be
{\widehat \P} \(t, t^{\pr} \) = - \int d \l \, \BA \,
S\(\P (t^{\pr}), \psi_{BA}^{\ast} (t^{\pr}, \l ) \)
\lab{whp}
\ee
We first observe that
$ \pa {\widehat \P} \(t, t^{\pr} \) / \pa t^{\pr}_n =0 $
due to eqs. \rf{potentialflo} and  \rf{bilide}.
Hence ${\widehat \P} \(t, t^{\pr} \)= {\widehat \P} \(t \)$ does
not dependent on the multi-time $t^{\pr} $.
Moreover, it is obvious from the definition \rf{whp} that
${\widehat \P} \(t \)$ is an eigenfunction possessing spectral representation
of the form \rf{spec} and, therefore, satisfying the conditions of Lemma
(\ref{lemma:spec}). Consequently, according to \rf{spec1}, we have:
\be
{\widehat \P} \(t \) = \int d \l\, \BA { 1\o \l} \BAAc{t^{\pr}}{\l}
{\widehat \P} \(t^{\pr}+ [\l^{-1}]\)
\lab{whpa}
\ee
Agreement between \rf{whp} and \rf{whpa} requires that their respective
$\l$-integrands may differ by at most a term proportional to the Hirota
$\l$-integrand in eq.\rf{bilide}. The latter implies the fulfillment of the
following identity:
\be
{ 1\o \l} \BAc {\widehat \P} \(t + [\l^{-1}]\)
= - S\(\P (t), \psi_{BA}^{\ast}(t,\l )\) + {\widehat A} \BAc
\lab{whpb}
\ee
where ${\widehat A}$ is some differential operator w.r.t.
$t= \( t_1 ,t_2 ,\ldots \)$.
Acting with $\pa_x$ on both sides of \rf{whpb} and using identity
\rf{xidentityc} together with \rf{whpa}, we conclude that 
${\widehat A}$ has to satisfy
\be
{\widehat \P} \(t \) - \P \(t \) = -{ \pa_x {\widehat A} \BAc \o \BAc}
\lab{whpc}
\ee
The left hand side of \rf{whpc} is a KP eigenfunction, independent of the
spectral parameter $\l$. This forces ${\widehat A}=0$ and
consequently  ${\widehat \P} \(t \) = \P \(t \) $.
This concludes the proof of Prop.\ref{proposition:propspec}. $\Box$

\begin{corollary}
Taking into account Prop.\ref{proposition:propspec},
eqs.\rf{fayx}--\rf{xidentityc} imply the following relations:
\br
S\(\psi_{BA}(t,\m ) ,\BAc\)&=& - { 1\o \l}  \psi_{BA} \(t + [\l^{-1}], \mu \) \BAc
=- \frac{\X \t (t)}{\t (t)}
\lab{X-S} \\
S\(\BA, \psi^{\ast}_{BA}(t,\m ) \)&=& 
{ 1\o \l} \BA \psi^{\ast}_{BA} \(t - [\l^{-1}], \mu \)  =
- \frac{{\widehat {\cal X}} \(\m, \l\) \t (t)}{\t (t)} +\d (\m,\l)
 \nonu \\
\phanta
\lab{S-X} \\
S \(\P (t),\BAc\) &=& - { 1\o \l} \BAc \P \( t + [\l^{-1}]\)
\lab{whpd} \\
S \(\BA ,\Psi (t)\)  &=&{ 1\o \l} \BA  \Psi \(t - [\l^{-1}]\)
\lab{whpe}
\er
where $\P, \Psi$ are arbitrary (adjoint-)eigenfunctions and
$S\(\cdot ,\cdot \)$ are the corresponding squared eigenfunction potentials.
Moreover, we also have the following double spectral density representation
for the SEP $S\(\P (t), \Psi (t)\) $:
\be
S\(\P (t), \Psi (t)\) =
- \int\!\int d\l\, d\m\, \vp^{\ast}(\l ) \vp (\m ) {1\o {\l}} \BAc
\psi_{BA} (t+[\l^{-1}],\m )
\lab{S-spec}
\ee
\label{corollary:whpc}
\end{corollary}

Taking into account  \rf{whpd}--\rf{whpe}, the spectral representations
\rf{spec1}--\rf{spec2} become:
\br
\P (t) &=& - \int d\l\,\BA\, S\(\P (t^\pr ), \psi_{BA}^{\ast}(t^\pr ,\l )\)
\lab{spec1-a}\\
\Psi (t) &=& \int d \l\,\BAc\, S\(\psi_{BA} (t^\pr ,\l ), \Psi (t^\pr )\)
\lab{spec2-a}
\er
\mark
Note that the expressions \rf{spec1-a}--\rf{spec2-a} applied for
(adjoint) BA functions yield:
\br
\BA &=& - \int d \m \, \psi_{BA} ( t, \m ) S \( \psi_{BA} ( t^{\pr}, \l ),
\psi_{BA}^{\ast} ( t^{\pr}, \m )\) \nonu\\
\BAc &=&\int d \m \, \psi_{BA}^{\ast}  ( t, \m ) S \( \psi_{BA}
( t^{\pr}, \m ), \psi_{BA}^{\ast} ( t^{\pr}, \l )\)
\lab{cauchy}
\er
which shows that the SEP
$S \( \psi_{BA} ( t^{\pr}, \l ), \psi_{BA}^{\ast} ( t^{\pr}, \m )\)$
can be identified with the Cauchy kernel for each fixed KP multi-time
$t^{\pr}$ (cf. also \ct{orlov2} and references therein, where the above SEP
was previously introduced in the context of Riemann factorization problem,
as well as \ct{carroll} for related discussion within the dispersionless 
KP hierarchy). 

\mark
Going back to the spectral representation eqs.\rf{spec1}--\rf{spec2}, valid
for any eigenfunction of the general KP hierarchy, we observe that they can be
rewritten as evolution equations w.r.t. the KP multi-time of the following
form:
\br
\P (t) = {\hat U} (t,t^\pr ) \P (t^\pr ) \qquad , \quad
{\hat U} (t,t^\pr ) \equiv \int d\l\,\BA{1\o\l}\BAAc{t^{\pr}}{\l}
e^{\sumi{1} {1\o l}\l^{-l}\pa/\pa t^\pr_l }
\lab{U-evol} \\
\Psi (t) = {\hat U}^{\ast}(t,t^\pr ) \Psi (t^\pr ) \qquad , \quad
{\hat U}^{\ast}(t,t^\pr ) \equiv \int d\l\,\BAc{1\o\l}\BAA{t^{\pr}}{\l}
e^{-\sumi{1} {1\o l}\l^{-l}\pa/\pa t^\pr_l}
\lab{U-evol-c}
\er
One can readily verify that:
\br
{\hat U}(t,t) = \one \quad , \quad {\hat U}^{-1}(t,t^\pr ) = {\hat U}(t^\pr ,t)
\quad , \quad
{\hat U}(t,t^\pr ) = {\hat U}(t,t^{\pr\pr})\, {\hat U}(t^{\pr\pr},t^\pr )
\lab{U-prop-1} \\
\partder{}{t_l}{\hat U}(t,t^\pr ) = L^{{l/r}}_{+} {\hat U}(t,t^\pr )
\quad , \quad
\partder{}{t^\pr_l}{\hat U}(t,t^\pr ) =
- {L^\pr}^{{l/r}}_{+} {\hat U}(t,t^\pr )
\lab{U-prop-2}
\er
{}From \rf{U-prop-1}--\rf{U-prop-2} we deduce that the evolution operator
${\hat U} (t,t^\pr )$ \rf{U-evol} can be formally written as a path-ordered
exponential:
\be
{\hat U} (t,t^\pr ) = P \exp \biggl\{ \sum_{l=1}^{\infty} \int_0^1 ds\,
\frac{d t_l}{ds}\, L^{{l/r}}_{+} \( t (s)\) \biggr\} \quad ; \quad
t_k (0) = t^\pr_k \;\; ,\;\; t_k (1) = t_k \;\; ,\; k=1,2,\ldots
\lab{U-path-int}
\ee
which precisely agrees with the formal solution of the differential
evolution eqs.\rf{eigenlax} for the KP eigenfunctions.
The r.h.s. of \rf{U-path-int} is independent of the particular path 
$\lcurl t_k (s) \rcurl$ connecting the points $t^\pr$ and $t$ in the space of 
KP multi-times due to the ``zero-curvature'' Zakharov-Shabat equations:
\be
\partder{}{t_k} L^{{l/r}}_{+} - \partder{}{t_l} L^{{k/r}}_{+} 
- \Sbr{L^{{k/r}}_{+}}{L^{{l/r}}_{+}} = 0
\lab{ZS}
\ee
Thus, our SEP method allowed us to find the explicit expression (r.h.s. of
the second
eq.\rf{U-evol}) for the formal path-ordered exponential \rf{U-path-int}.

Now, it is worthwhile to observe that we can revert the logic of our procedure 
above, {\sl i.e.}, instead
of starting with Hirota bilinear identity \rf{bilide} (or, equivalently, with
Fay identity \rf{fayid}) as defining the KP hierarchy and deriving from them 
the spectral representation formalism \rf{spec1}--\rf{spec2}
(or \rf{spec1-a}--\rf{spec2-a}) for KP eigenfunctions, we can take the spectral
representation eqs.\rf{spec1-a}--\rf{spec2-a} as the basic equations defining
the KP hierarchy. Namely, we have the following simple:
\begin{proposition}
Consider a pair of functions $\psi (t,\l ), \psi^{\ast}(t,\l )$ of the
multi-time $\( t_1 ,t_2 ,\ldots\)$ and the spectral parameter $\l$ of the form
$\, \psi^{(\ast )}(t,\l ) = e^{\pm \xi (t,\l )} \sum_{j=0}^{\infty}
w^{(\ast )}_j (t) \l^{-j}$ with $w^{(\ast )}_0 =1$ and $\xi (t,\l )$ as in
\rf{xidef}. Let us assume that $\psi^{(\ast )}(t,\l )$ obey the following 
spectral identities:
\be
\psi (t,\l ) = - \int d\m\, \psi (t,\m ) \, S(t^\pr ;\l,\m ) \quad , \quad
\psi^{\ast}(t,\l ) = \int d\m\, \psi^{\ast} (t,\m )\, S(t^\pr ;\m,\l )
\lab{cauchy-0}
\ee
for two arbitrary multi-times $t$ and $t^\pr$, where by definition the 
function $S(t;\l,\m )$ is such that 
$\partder{}{t_1} S(t;\l,\m ) = \psi (t,\l ) \,\psi^{\ast}(t,\m )$.
Then, eqs.\rf{cauchy-0} are equivalent to Hirota bilinear identity \rf{bilide}
and, accordingly, $\psi^{(\ast )}(t,\l )$ become (adjoint) BA functions of
the associated KP hierarchy.
\label{proposition:revert}
\end{proposition}
To see that eqs.\rf{cauchy-0} imply Hirota identity \rf{bilide}, it is enough
to differentiate both sides of \rf{cauchy-0} w.r.t. $t_1^\pr$: 
$0 = \pa \psi (t,\l ) / \pa t^\pr_1 = - \psi (t^\pr ,\l )
\int d\m\, \psi (t,\m )\,\psi^{\ast}(t^\pr ,\m )$.
The proof of the inverse statement of the equivalence, namely, that Hirota 
bilinear identity \rf{bilide} imply the spectral representation 
eqs.\rf{cauchy-0}, is contained in the proof of
Prop.\ref{proposition:propspec} above. 
\lskip
Using \rf{psi-main}--\rf{psi-mainc}, eqs.\rf{whpd}-\rf{whpe} can be rewritten
as:
\br
{ \t \(t + [\l^{-1}]\) \P \(t + [\l^{-1}]\) \o \l \t (t) } e^{-\xi (t, \l)}
&=&- S \( \P (t), \BAc \)
\lab{whpdd} \\
{ \t \(t - [\l^{-1}]\) \Psi \(t - [\l^{-1}]\) \o \l \t (t) } e^{\xi (t, \l)}
&=&
S \(\BA , \Psi (t) \)
\lab{whpee}
\er

\mark
Spectral representations for eigenfunctions \rf{spec1-a}--\rf{spec2-a} as well
as identities \rf{whpdd}--\rf{whpee} were obtained in a similar form in 
\ct{z-cheng} for the particular case of constrained \cKPrm hierarchies. 
Let us specifically emphasize, that all main equations of the present SEP method 
\rf{spec}--\rf{spec2}, \rf{X-S}--\rf{spec2-a} and
\rf{whpdd}--\rf{whpee}, derived above, are valid within the general 
unconstrained KP hierarchy.

Acting with space derivative $\pa_x$ on both sides of \rf{whpdd}-\rf{whpee}
and shifting the KP time arguments, we get:
\be
\frac{\P\bigl( t-[\l^{-1}]\bigr)}{\P (t)} - 1 + \l^{-1} \pa \ln \P (t) =
\l^{-1} \pa \ln \frac{\t \bigl( t-[\l^{-1}]\bigr)}{\t (t)}
\lab{tau-Phi-id}
\ee
\be
\frac{\Psi\bigl( t+[\l^{-1}]\bigr)}{\Psi (t)} - 1 - \l^{-1} \pa \ln \Psi (t) =
- \l^{-1} \pa \ln \frac{\t \bigl( t+[\l^{-1}]\bigr)}{\t (t)}
\lab{tau-Psi-id}
\ee
which were obtained in \ct{noak} by studying the way the $\t$-function
transforms under \DB transformations.
Taking into consideration that:
\be
- \l+ \l  \frac{\P\bigl( t-[\l^{-1}]\bigr)}{\P (t)} +\pa \ln \P (t)
 = \sumi{n=2}  { p_n ( - \lb \pa \rb ) \P (t)\over \l^{n-1} \P (t)}
\lab{212a}
\ee
with $p_n (\cdot )$ being the Schur polynomials \rf{Schur},
we find that eq.\rf{tau-Phi-id} is a generating 
equation for the following set of equations upon expanding in powers of
$\l^{-1}$ :
\br
p_n ( - \lb \pa \rb ) \P (t) &= & v_n (t) \P (t) \quad ;\quad n \geq 2
\lab{sato2} \\
v_n (t) &\equiv& p_{n-1} ( - \lb \pa \rb ) \,  \pa \ln \t (t) \nonu
\er
Note that $v_n (t)$ are coefficients in the $\l$-expansion of the
generating function $v (t, \l )$ \ct{tak2} :
\be
v (t, \l ) = \sumi{n=1} v_{n+1} \l^n \equiv  \pa_x \ln \BA - \l =
{\widehat \Delta}_\l \pa_x \ln \t (t)
\lab{vtl}
\ee
where in obtaining the last equality we again used
eqs.\rf{psi-main}--\rf{psi-mainc} and notation \rf{deltal}.
We will later need a slight generalization of \rf{vtl} :
\be
v^{(l)} (t, \l ) = \sumi{n=1} \s^{(l)}_{n} (t) \l^n \equiv  
\partder{}{t_l} \ln \BA - \l^l =
{\widehat \Delta}_\l \partder{}{t_l} \ln \t (t) \quad ; \quad l \geq 1
\lab{vtl-l}
\ee
Clearly $\s^{(l)}_{n}(t) = p_{n} ( - \lb \pa \rb ) \,\pa/\pa t_l \ln \t$ and
$v_n (t) = \s^{(1)}_{n-1}(t) \, ,\, n\geq 2$.
The coefficients $\s^{(l)}_{n}$ enter the basic identity for the KP Lax
operator \rf{lax-op} :
\be
\( L^{l/r} \)_{+} = L^{l/r} + \sumi{n=1}\s^{(l)}_{n} L^{-n/r}
\lab{basickp}
\ee

\mark
Eqs.\rf{sato2} are, clearly, valid for an arbitrary eigenfunction $\P$ of the
full KP hierarchy. 
On the other hand, in ref.\ct{KP} (see also \ct{chengs}) eqs.\rf{sato2}
were presented for the special case of $\P = \BA$ as relations equivalent
to the standard KP evolution equations $\pa \BA/ \pa t_n = (L^{n/r})_{+} \BA$.
In fact, as shown in \ct{tak2}, plugging the BA wave function 
$\P (t) = \psi_{BA} (t, \mu )$ into eq.\rf{tau-Phi-id}
one easily recovers the differential Fay identity \rf{dfayid}.  

We now define the ``ghost'' symmetry flows generated  by the SEP
\ct{oevela,chengs,noak}.
Let $\pa_{\a}$ be a vector field, whose action on the Lax operator $L$ and,
accordingly, on the dressing
operator $W$, is induced by a set of (adjoint) eigenfunctions
$\P_a, \Psi_a, \, a \in \{ \a \}$ through:
\be
\pa_\a L \equiv \Sbr{\sum_{a \in \{ \a \}} \Phi_a D^{-1} \Psi_a}{L}
\quad ; \quad
\pa_{\a} {W} \equiv \biggl( \sum_{a \in \{ \a \}}
 \Phi_a D^{-1} \Psi_a \biggr) W
\lab{ghostw}
\ee
As shown in \ct{oevela}, the corresponding action of the above ``ghost''
flows on the (adjoint) eigenfunctions $\P$, $\Psi$: 
\be
\pa_{\a} \P  =  \sum_{a \in \{ \a \}}  \P_a S \(\P, \Psi_a \)
\quad ;\quad \pa_{\a} \Psi = \sum_{a \in \{ \a \}} S \(\P_a, \Psi \)\Psi_a
\lab{paaonp}
\ee
is compatible with the isospectral evolutions of
$\P$, $\Psi$.
Furthermore, it is easy to see that
\be
\pa_{\a} S \(\P, \Psi \)  =   \sum_{a \in \{ \a \}}
S \(\P, \Psi_a \) S \(\P_a, \Psi \)
\lab{paaons}
\ee
is compatible with eq.\rf{paaonp}

If 
$\pa_{\b} {W} \equiv \Bigl( \sum_{b \in \{\b \}} \P_b D^{-1} \Psi_b\Bigr) W$
defines some other ``ghost'' flow and both flows $\pa_{\a}$ and
${\pa_{\b}}$ satisfy \rf{paaonp}, then:
\be
\sbr{\pa_{\a}}{\pa_{\b}} W =0
\lab{ghostcom}
\ee
as follows from the technical identity \rf{x1x2}.
Equations \rf{paaonp} and \rf{ghostcom} can be compactly expressed
by an identity
$ \sbr{\pa_{\a}-\sum_{a \in \{\a \}} \P_a D^{-1} \Psi_a}
{\pa_{\b}-  \sum_{b \in \{\b \}} \P_b D^{-1} \Psi_b}=0$
\ct{orlov1,oevela}.

Define now
$Y (\l,\m) \equiv \psi_{BA} \(t, \m \) D^{-1} \psi^{\ast}_{BA}(t,\l )$
(cf. ref.\ct{Dickey-addsym}) to be pseudo-differential operator 
inducing a ghost-flow ~$ \pa_{(\l,\m)} W \equiv Y (\l,\m) W$
according to \rf{ghostw}. In this case the ``SEP'' symmetry flow
is generated by an infinite combination of $\Win1$ algebra generators
\ct{Dickey-addsym}. Then, according to eq.\rf{paaonp} the action of this
flow on the BA wave function is given by:
\be
{\widehat Y}(\l,\m) \bigl(\psi_{BA}(t,z)\bigr) \equiv
\pa_{(\l,\m)} \bigl(\psi_{BA}(t, z) \bigr) = \psi_{BA} \(t, \m\)
S\(\psi_{BA}(t,z ) , \psi^{\ast}_{BA}(t,\l ) \)
\lab{Ydef}
\ee
Further, let us also define the action of the vertex operator $\X$ on
the BA function $\psi_{BA}(t,z)$ as generated by its action
(as a vector field) on the ratio of $\t$-functions entering \rf{psi-main} :
\be
\X \psi_{BA} \(t, z \) =
e^{\xi (t,z)} { \t (t)\X \t (t-[z^{-1}]) - \t (t - [z^{-1}])
\X \t (t) \o \t^2 (t) }
\lab{xonpsi-def}
\ee
The latter, upon using the shift-difference operator \rf{deltal},
can be written as:
\be
\X \psi_{BA} \(t, z \) =
\psi_{BA} \(t, z \)  {\widehat \D}_{z} {\X \t (t) \o \t (t) }
\lab{xonpsiz}
\ee
Let us stress that, according to \rf{Ydef}--\rf{xonpsiz},
${\widehat Y}(\l,\m)$ acts on the BA function as a standard
pseudo-differential operator, whereas $\X$ acts on it as a shift-difference
operator.

Now, the above results allow us to give a simple straightforward proof
of the following version of the Adler-Shiota-van-Moerbeke proposition
\ct{ASvM,Dickey-addsym}.
It provides the connection between the form of
the {\em non-isospectral (``additional'') symmetries} of KP hierarchies
acting on the Lax operators and BA functions \ct{Orlovetal}, on one hand, and
their respective form when acting on KP $\t$-functions, on the other hand.
\begin{corollary}
With definitions \rf{Ydef} and \rf{xonpsi-def} it holds:
\be
\X \psi_{BA}(t,z) = {\widehat Y}(\l,\m) \bigl(\psi_{BA}(t,z)\bigr)
\lab{x=y}
\ee
\label{corollary:x=y}
\end{corollary}
\begin{proof}
Indeed, applying \rf{xidentityb} and Lemma \ref{lemma:deltaonpsi}
to the r.h.s. of \rf{xonpsiz}, the latter equation can be rewritten as:
\be
\X \psi_{BA} \(t, z \) = {1\o z} \psi_{BA} \(t, z \) \psi_{BA} \(t, \m \)
\psi^{\ast}_{BA}(t-[z^{-1}],\l ) =
{\widehat Y} (\l,\m) \(\psi_{BA} \(t, z \) \)
\lab{x=ya}
\ee
where in order to arrive at the last equality use was made of \rf{S-X}.
\end{proof}

In the literature one often comes across the vertex operator defined as
$ {\widehat X} (\l, \m) \equiv\;\\
: \exp \({\hat \th}(\l ) - {\hat \th}(\m )\):\; = (\l -\m) \X$.
In such a notation the expression \rf{x=y} becomes
${\widehat X}(\l,\m) = (\l -\m) {\widehat Y}(\l, \m)$ as in 
\ct{ASvM,Dickey-addsym}.

We conclude this section by proving the following important property of SEP:
\begin{lemma}
Under shift of the KP times, the squared eigenfunction potential obeys:
\br
S\(\P (t-[\l^{-1}]), \Psi (t-[\l^{-1}])\) - S\(\P (t), \Psi (t)\) =
- {1\o {\l}} \P (t) \Psi (t-[\l^{-1}])
\lab{S-id-1}  \\
S\(\P (t+[\l^{-1}]), \Psi (t+[\l^{-1}])\) - S\(\P (t), \Psi (t)\) =
{1\o {\l}} \P (t+[\l^{-1}]) \Psi (t)
\lab{S-id-2}
\er
\label{lemma:S-lemma}
\end{lemma}
\begin{proof}
According to \rf{S-spec} and \rf{X-S} :
\be
{\widehat \D}_{z} S\(\P (t), \Psi (t)\) =
\int\!\int d\l\, d\m\, \p^{\ast}(\l ) \p (\m )
{\widehat \D}_{z} S \( \psi_{BA} ( t, \l ),\psi_{BA}^{\ast} ( t, \m )\)
\lab{S-speca}
\ee
while from eq.\rf{delpsi2} we find that:
\be
{\widehat \D}_{z} S \( \psi_{BA} ( t, \l ),\psi_{BA}^{\ast} ( t, \m )\) =
- {1 \o z} \psi_{BA} (t,\l) \psi_{BA}^{\ast} (t-[z^{-1}],\m)
\lab{S-specb}
\ee
Inserting the above identity back in \rf{S-speca} gives \rf{S-id-1}.
\end{proof}
After expanding identities \rf{S-id-1} and \rf{S-id-2}
in power series w.r.t. $\l$ we obtain:
\br
p_s (-[\pa]) S\biggl(\P (t),\Psi (t)\biggr) &=&
- \P (t) \, p_{s-1}(-[\pa]) \Psi (t)  \nonu  \\
p_s ([\pa]) S\biggl(\P (t),\Psi (t)\biggr) &=&
 \Psi (t)\,  p_{s-1}([\pa]) \P (t) \quad , \quad \;\; s=1,2,\ldots
\lab{S-id-comp}
\er
where $p_s (\cdot )$ are the standard Schur polynomials \rf{Schur}.
\sect{Constraints on ${\sf cKP}_{r,m}$ Tau-Functions.
Grassmannian Interpretation}
\label{section:grassmann}
{}From now on we concentrate on studying the class of constrained \cKPrm
hierarchies for which we have: 
\be
\( L_{r,m}\)_{-} = \sum_{a=1}^m \P_a D^{-1} \Psi_a
\lab{ckprm}
\ee
according to eq.\rf{f-5}.
We first note that the \cKPrm BA function satisfies, according to \rf{ckprm}, 
the following spectral equation:
\be
L_{r,m} \BA = \l^r \BA = (L_{r,m})_{+} \BA + \sum_{a=1}^m \P_a (t)
\, S \( \BA , \Psi_a (t)\)
\lab{grassckp}
\ee
Due to eq.\rf{whpe}, the latter can be cast in the following form:
\br
\l^r \BA = (L_{r,m})_{+} \BA + \sum_{a=1}^m { 1\o \l} \P_a (t)
  \Psi_a \(t - [\l^{-1}]\) \BA   \nonu  \\
= (L_{r,m})_{+} \BA - \sum_{a=1}^m \Bigl\lb
S\(\P_a (t-[\l^{-1}]),\Psi_a (t-[\l^{-1}])\) - 
S\(\P_a (t),\Psi_a (t)\)\Bigr\rb\, \BA
\lab{grassckpa}
\er
where the second equality in \rf{grassckpa} follows from \rf{S-id-1}.
Recalling relation \rf{vtl-l} we find that
$\pa \t (t)/ \pa t_r = \sum_{a=1}^m S\(\P_a (t),\Psi_a (t)\) \t (t)$.
Similarly, using the spectral identity $L_{r,m}^n \BA = \l^{rn} \BA$
and taking into account relation \rf{lkminus} we obtain the following set of
differential equations for the \cKPrm $\t$-function:
\be
\partder{}{t_{rn}} \t (t) = \Biggl\lb \sum_{a=1}^m \sum_{i=0}^{n-1}
S\( L^{n-1-i}(\P_a ), {L^{\ast}}^i (\Psi_a )\) \Biggr\rb \t (t)
\lab{tau-t-n}
\ee
Using the differential Fay identity \rf{dfayid}, eqs.\rf{tau-t-n} can be
equivalently written in the form:
\be
\lcurl \partder{}{t_{rn}} - \Biggl\lb \partder{}{t_{rn}}\, ,\,
\int\!\int d\l\, d\m\, \(\l^r - \m^r\)^{-1}
\sum_{a=1}^m \vp^{\ast}_a (\l ) \vp_a (\m ) \X \Biggr\rb \rcurl \t (t) = 0
\lab{tau-t-n-X}
\ee
where $\vp^{(\ast )}_a (\l )$ are the ``spectral densities'' of the
(adjoint) eigenfunctions $\P_a (t), \Psi_a (t)$ entering the
pseudo-differential part of the \cKPrm Lax operator \rf{f-5}, and also we
have used the identity:
\be
\sbr{\partder{}{t_l}}{\X} = \(\m^l - \l^l\) \X
\lab{t-X-id}
\ee
Thus we arrive at the following statement providing an alternative
definition of \cKPrm hierarchies intrinsically in terms of $\t$-functions:
\begin{proposition}
Reduction of the full KP hierarchy \rf{lax-op} to the \cKPrm hierarchy in
terms of Lax operators \rf{f-5} is equivalent to imposing eqs.\rf{tau-t-n-X}
as constraints on the pertinent $\t$-functions, where 
$\vp^{(\ast )}_a (\l )$ are ``spectral densities'' of KP (adjoint)
eigenfunctions given as in eqs.\rf{specdens}.
\label{proposition:cKP-tau}
\end{proposition}
Let us now translate eq.\rf{tau-t-n-X} into the language of universal Sato
Grassmannian ${\cG}r$ \ct{KP,OSTT}. Consider the hyperplane $\cW \in {\cG}r$
defined through a linear basis of Laurent series
$\lcurl f_k (\l ) \rcurl$ in $\l$ in terms of the BA function
as generating function $F(t,\l )$ :
\br
\cW \equiv {\rm span } \langle f_1 (\l )\!\!\!&,&\!\!\! f_2 (\l )\,,
\ldots \rangle \nonu \\
f_k (\l ) = \partderh{}{x}{k} F(t,\l ) \bgv_{x= t_2 =t_3 = \ldots =0}
\quad &,&\quad  F(t,\l ) = \BA
\lab{wgras}
\er
In case of the standard $r$-th KdV reduction, where the corresponding
Lax operator $\cL = D + \sumi{1} u_i D^{-i}$ satisfies $\cL^r = \cL^r_{+}$,
the latter constraint translates to the Grassmannian language as
$\l^r \cW \subset \cW\,$ \ct{Miwa-Jimbo}.

Our aim now is to express the \cKPrm constraint \rf{ckprm} (cf. \rf{f-5}) 
in the Grassmannian setting. 
We find from \rf{grassckpa} that the generating function
$F^{\pr}(t,\l )$ :
\be
F^{\pr}(t,\l ) \equiv \Bigl\lb \l^r + \sum_{a=1}^m  \sumi{n=1}
{ p_n ( - \lb \pa \rb ) S \(\P_a (t),\Psi_a (t) \)\over \l^{n}} \Bigr\rb \BA
= (L_{r,m})_{+} \BA
\lab{inw}
\ee
defines via \rf{wgras} a point $\cW^\pr$ of Sato Grassmannian ${\cG}r$ :
\be
\cW^\pr = {\rm span } \langle F^{\pr}(0,\l ), \pa_x F^{\pr}(0,\l ),
\pa^2_x F^{\pr}(0,\l ), \ldots \rangle
\lab{wgras-a}
\ee
which coincides, because of the second equality in \rf{inw}, with the
original point $\cW$ defined through $F(t,\l ) = \BA\,$ \rf{wgras}.
Thus, we have\foot{For a different criteria characterizing \cKPrm hierarchies
within the Sato Grassmannian framework, see refs.\ct{zhang,vandeleur}.} :
\begin{proposition}
Let $S\(\P_a (t),\Psi_a (t)\)$ , $a=1,\ldots ,{\rm m}$ , be ${\rm m}$
squared eigenfunction potentials \rf{potentialflo} where $\P_a ,\Psi_a$ are
(adjoint-)eigenfunctions of the general KP hierarchy \rf{lax-op}. Then,
the reduction of \rf{lax-op} to the \cKPrm hierarchy \rf{f-5} can be
equivalently expressed as a restriction of ${\cG}r$ to a subset whose
points (hyperplanes) $\cW$ \rf{wgras} are subject to the following constraint:
\be
\llb \l^r + \sum_{a=1}^m \, {\widehat \D}_{\l} \, S \(\P_a (t),\Psi_a (t)\)
\rrb \cW \subset \cW
\lab{inww}
\ee
with ${\widehat \D}_{\l}$ as in \rf{deltal} and $S \(\P_a (t),\Psi_a (t)\)$
being given by \rf{S-spec} in terms of the generating function \rf{wgras} of
$\cW$.
\label{proposition:ckp-grass}
\end{proposition}

\sect{Non-Isospectral Virasoro Symmetry for \cKPro $\t$-func\-tions}
\label{section:virasoro}
The conventional formulation of additional non-isospectral symmetries 
for the full KP integrable hierarchy \ct{Orlovetal,cortona}
is not compatible with the reduction of the latter to the important 
class of constrained \cKPrm integrable models. In refs.\ct{addsym,noak} we
solved explicitly the problem of compatibility of the Virasoro part
of non-isospectral symmetries with the underlying constraints of \cKPrm
hierarchies within the pseudo-differential Lax operator framework.
Our construction in \ct{addsym,noak} involves an appropriate modification 
of the standard non-isospectral symmetry flows, acting on the space of
\cKPrm Lax operators, by adding a set of additional ``ghost symmetry'' flows
(of the type appearing in eq.\rf{ghostw}). In this section, we derive the
explicit form of the action of the correct modified Virasoro non-isospectral
symmetries as flows on the space of \cKPrm $\t$-functions. Note that the
corresponding result for the full unconstrained KP hierarchy has been
previously obtained in \ct{ASvM,Dickey-addsym,cortona}.

To this end, let us first recall that the standard
{\em additional (non-isospectral) symmetries}
\ct{Orlovetal,cortona} are defined as vector fields on the space of general
KP Lax operators \rf{lax-op} or, alternatively, on the dressing
operators \rf{dress-1}, through their flows as follows:
\be
{\bar \pa}_{k,n} L = - \Sbr{\( M^n L^k\)_{-}}{L} =
\Sbr{\( M^n L^k\)_{+}}{L} + n M^{n-1} L^k  \;\,;  \;\;\;\,
{\bar \pa}_{k,n} W = - \( M^n L^k\)_{-} W
\lab{add-symm-L}
\ee
Here $M$ is a pseudo-differential operator ``canonically conjugated'' to 
$L$ such that:
\be
\Sbr{L}{M} = \one \quad , \quad
\partder{}{t_l} M = \Sbr{L^{l/r}_{+}}{M} 
\lab{L-M}
\ee
Within the Sato-Wilson dressing operator formalism, the $M$-operator can be 
expressed in terms of dressing of the ``bare'' $M^{(0)}$ operator:
\be
M^{(0)} = \sum_{l \geq 1} \frac{l}{r} t_l D^{l-r} =
X_{(r)} + \sum_{l \geq 1} \frac{l+r}{r} t_{r+l} D^l \quad ; \quad
X_{(r)} \equiv \sum_{l=1}^{r} \frac{l}{r} t_l D^{l-r}
\lab{M-0}
\ee
conjugated to the ``bare'' Lax operator $L^{(0)} = D^r$.

The additional symmetry flows \rf{add-symm-L} commute with the usual KP 
hierarchy isospectral flows given in \rf{lax-eq}. 
However, they do not commute among themselves,
instead they form a centerless $\Win1$ algebra (see e.g. \ct{cortona}).
One finds that the Lie algebra of operators ${\bar \pa}_{k,n}$
is isomorphic to the Lie algebra generated by $- z^k (\pa/\pa z)^n$.
Especially for $n=1$ this becomes an isomorphism to the centerless Virasoro
algebra ${\bar \pa}_{k,1} \sim - \cL_{k-1}$, with $\sbr{\cL_l}{\cL_k} = (l-k)
\cL_{l+k}$.

As demonstrated in \ct{addsym,noak}, the conventional non-isospectral flows
\rf{add-symm-L} do {\em not} preserve the space of \cKPrm Lax operators
given by \rf{f-5}. In particular, for the Virasoro non-isospectral symmetry
algebra the transformed Lax operator ${\bar \pa}_{k,1} L$ belongs to a 
{\em different} class of constrained KP hierarchies -- ${\sf cKP}_{r,m(k-1)}$
(when $k \geq 3$). The solution to this problem is provided by the
following \ct{addsym,noak}:
\begin{proposition}
The correct non-isospectral symmetry flows for the \cKPrm hierarchies 
\rf{f-5}, spanning the Virasoro algebra, are given by:
\be
\pa^{\ast}_k\, L  \;\equiv\;  \sbr{- \( M L^k \)_{-} + X^{(1)}_{k-1}}{L}
\lab{pasta}
\ee
{\sl i.e.}, with the following isomorphism
$\cL_{k-1} \sim -\( M L^k \)_{-}  + X^{(1)}_{k-1}$, where $X^{(1)}_{k-1}$ are
ghost-symmetry generating operators (cf. \rf{ghostw}) defined as:
\be
X^{(1)}_k = \sum_{i=1}^m \sum_{j=0}^{k-1} \(j - \h (k-1)\) L^{k-1-j}
(\Phi_i) D^{-1} \( L^{\ast}\)^{j} (\Psi_i) \quad ;\quad k \geq 1 
\lab{xkb}
\ee
\label{proposition:mainprop}
\end{proposition}

Since (auto-)\DB transformations of \cKPrm hierarchies (see next section)
play a fundamental r\^{o}le for finding exact solutions, as well as in
establishing the link between \cKPrm integrable models and (multi-)matrix
models, it is natural to impose the additional condition of commutativity of the 
non-isospectral symmetries with the \DB transformations. The latter condition
was shown in refs.\ct{addsym,noak} to be satisfied only by the subclass
\cKPro of constrained KP hierarchies (it is precisely \cKPro hierarchies
which provide the integrability structure of discrete multi-matrix models
\ct{noak}). Therefore, in the rest of this section we restrict our attention
to \cKPro models.

Consider the modified non-isospectral Virasoro symmetry flows \rf{pasta} 
acting on the dressing operator of \cKPro hierarchy:
\be
\pa^{\ast}_k W = - \( ML^k\)_{-} W + X^{(1)}_{k-1} W
\lab{Vir-W}
\ee
Taking the operator residuum on both parts of \rf{Vir-W} we obtain:
\be
\pa^{\ast}_k \t (t) = {1\o {2r}}{\widehat W}^{(2)}_{r(k-1)} \t (t)
+ \Biggl\lb \sum_{j=0}^{k-2} \(\h (k-2) -j\)
S\(L^{k-2-j}(\P ), {L^{\ast}}^j (\Psi )\) \Biggr\rb \t (t)
\lab{Vir-tau}
\ee
In deriving eq.\rf{Vir-tau} we used the expression for $X^{(1)}_{k-1}$
\rf{xkb} together with the differential Fay identity \rf{dfayid} as well as:
\br
&&\Res\( M^l L^k\) = Res_{\l} \llb \( M^l L^k\) \(\BA\) \,
\BAc \rrb \nonu \\
&=& {1\o {r^l}} Res_{\l} \( \l^{kr -l(r-1)}
\partderh{}{\l}{l} \BA \; \BAc \) \lab{ASvM-r} \\
\eq  - \pa_x \llb {1\o {\t(t)}} Res_{\l} \( {1\o {r^l}} \m^{kr -l(r-1)}
\partderh{}{\m}{l} \X \bv_{\m =\l}\) \t (t) \rrb =
\pa_x \({1\o {r^l}(l+1)} \frac{{\widehat W}^{(l+1)}_{(k-l)r} \t (t)}{\t (t)}\)
\nonu
\er
In the chain of the identities in \rf{ASvM-r} we took into account
Dickey's formula for $\( M^l L^k\)_{-}$ \ct{Dickey-addsym}
(first equality in \rf{ASvM-r}), eq.\rf{fayx} (third equality in
\rf{ASvM-r}), and formula \rf{X-expand} for $\X$ to arrive at the last
equality above. 
The Virasoro operator in the 
first term on the r.h.s. of \rf{Vir-tau} comes from
the standard Orlov-Schulman non-isospectral symmetry flow and reads
explicitly (for $k \geq -1$) :
\be
{\widehat W}^{(2)}_{k} = 2 \sum_{l \geq 1} l t_l \partder{}{t_{l+k}}
- (k+1)\partder{}{t_k} + \sum_{l=1}^{k-1} \partderm{}{t_l}{t_{k-l}}
\lab{Vir-standard}
\ee
We now express the second additional ``ghost-flow'' term on the r.h.s. of
\rf{Vir-tau} as differential operator acting on $\t (t)$ of a form similar to \rf{Vir-standard}. The starting point are the differential equations
\rf{tau-t-n} obeyed by the \cKPro $\t$-function,
wherefrom we get for the second-order derivatives:
\br
\frac{1}{\t (t)}\partderm{\t (t)}{t_{rl}}{t_{rn}} &=&
\sum_{i=0}^{n-1}\llb S\( L^{n+l-1-i}(\P ), {L^{\ast}}^i (\Psi )\)
- S\( L^{n-1-i}(\P ), {L^{\ast}}^{i+l} (\Psi )\) \rrb  \nonu \\
&+& \sum_{i=0}^{n-1} \sum_{j=0}^{l-1}
\llb S\( L^{n-1-i}(\P ), {L^{\ast}}^i (\Psi )\)
S\( L^{l-1-j}(\P ), {L^{\ast}}^j (\Psi )\) \right.  \nonu \\
&-& \left. S\( L^{n-1-i}(\P ), {L^{\ast}}^j (\Psi )\)
S\( L^{l-1-j}(\P ), {L^{\ast}}^i (\Psi )\) \rrb
\lab{tau-t-ln}
\er
In obtaining relation \rf{tau-t-ln} we made use of the following Lemma:
\begin{lemma}
The relation: 
\br
\partder{}{t_{nr}} S (f,g) &=& S \( L^n (f), g\) - S \( f, {L^{\ast}}^n (g)\)
\nonu\\
&-& \sum_{i=0}^{n-1} S \( L^{n-1-i} (\P), g\) S \( f,{L^{\ast}}^i (\Psi) \)
\lab{panrS}
\er
holds for $f$ an eigenfunction and $g$ an adjoint eigenfunction of the Lax 
operator $L \equiv L_{r,1}= L_{+} + \P D^{-1} \Psi \,$ belonging to 
the \cKPro hierarchy.
\label{lemma:panrsfg}
\end{lemma}
\begin{proof}
We are going to show that
$\pa/\pa t_{nr}\, S (f,g) = {\rm Res} \( D^{-1} g (L^{n})_{+} f D^{-1} \)$
is equal to the right hand side of eq.\rf{panrS} (up to a constant).
We first apply $\pa/\pa t_{mr} $ on the left hand side of eq.\rf{panrS}.
This yields 
\br
\partder{}{t_{mr}}\partder{}{t_{nr}} S (f,g) &=& 
\partder{}{t_{nr}}\partder{}{t_{mr}} S (f,g) =
-{\rm Res} \( D^{-1} {(L^{\ast})}^n_{+} (g) L^{m} f D^{-1} \)
\nonu\\
&+&{\rm Res} \( D^{-1} g L^{m} (L)^n_{+} (f) D^{-1} \)
\lab{panrS1}
\er
After making the substitutions:
\br
(L)^n_{+} (f)&=&  L^n (f) - \sum_{i=0}^{n-1} L^{n-1-i} (\P)
S \( f,{L^{\ast}}^i (\Psi) \) \lab{panrSa}\\
{(L^{\ast})}^n_{+} (g) &=& {L^{\ast}}^n (g) +\sum_{i=0}^{n-1}
{L^{\ast}}^i (\Psi) S \( L^{n-1-i} (\P), g\)
\lab{panrSb}
\er
where use was made of \rf{lkminus},
we obtain agreement with the result of applying $\pa/\pa t_{mr}$
on the right hand side of eq.\rf{panrS} and using eq.\rf{potentialflo} 
as well as Lemma \ref{lemma:associati}
\end{proof}

Using eqs.\rf{tau-t-n},\rf{tau-t-ln} we obtain:
\be
\sum_{j=0}^{k-2} \( \h (k-2) -j\)
S\( L^{k-2-j}(\P ), {L^{\ast}}^j (\Psi )\) =
{1\o {2 \t (t)}} \sum_{l=1}^{k-2} \partderm{\t}{t_{rl}}{t_{r(k-1-l)}}
\lab{addflow-t}
\ee
Collecting \rf{Vir-standard} and \rf{addflow-t}, the final form of the
\cKPro non-isospectral Virasoro symmetry flows reads:
\br
\pa^{\ast}_k \t (t) &=&
\llb {1\o r}\sum_{l \geq 1} l t_l \partder{}{t_{l+r(k-1)}} -
\frac{r(k-1)+1}{2r}\partder{}{t_{r(k-1)}} \right.  \nonu  \\
 &+& \left. {1\o {2r}} \sum_{l=1}^{r(k-1)-1} \partderm{}{t_l}{t_{r(k-1)-l}} +
\h \sum_{l=1}^{k-2} \partderm{}{t_{rl}}{t_{r(k-1-l)}} \rrb \t (t)
\lab{pasta-Vir-r}
\er
In particular, for ${\sf cKP}_{1,1}$ hierarchies the Virasoro
non-isospectral symmetry takes the form (eq.\rf{pasta-Vir-r} for $r=1$) :
\be
\pa^{\ast}_k \t (t) =
\llb \sum_{l \geq 1} l t_l \partder{}{t_{l+k-1}} - {k\o 2}\partder{}{t_{k-1}}
+ \sum_{l=1}^{k-2} \partderm{}{t_{l}}{t_{k-1-l}} \rrb \t (t)
\lab{pasta-Vir-1}
\ee

Concluding this section it is instructive to point out the relation of 
\rf{pasta-Vir-1} with the so called Virasoro constraints in conventional 
discrete matrix models \ct{Morozov-MM,Morozov-Usp} spanning the Borel 
subalgebra of the Virasoro algebra:
\br
\cL^{(N)}_s Z_N &=& 0 \qquad ,\quad s \geq -1
\lab{Vir-Z-N} \\
\cL_s^{(N)} &=& \sumi{k=1} k t_k \partder{}{t_{k+s}} + 2 N \partder{}{t_s} +
\sum_{k=1}^{s-1} \partder{}{t_k}\partder{}{t_{s-k}} \quad , \quad s \geq 1
\lab{lsn1}  \\
\cL_{0}^{(N)} &=& \sumi{k=1} k t_k \partder{}{t_k} + N^2 \quad \; ; \; \quad
\cL_{-1}^{(N)} = \sumi{k=2} k t_k \partder{}{t_{k-1}} + N t_1
\lab{lsn2}
\er
Here $Z_N$ denotes the one-matrix model partition function with $N$ indicating
the size of the corresponding (Hermitian) random matrix. On one hand, it
can be identified as $\t$-function of the semi-infinite one-dimensional Toda
lattice model \ct{Morozov-MM}. On the other hand, from the point of view
of continuum integrable systems it was shown in \ct{avoda,noak} to be
$Z_N = \t^{(N,0)}$, {\sl i.e.}, $N$-th member of the \DB orbit on the subspace
of ${\sf cKP}_{1,1}$
$\t$-functions starting from the ``free'' initial $\t^{(0,0)} = 1$ 
(see next section for more details about \DB orbits on constrained KP 
hierarchies). Comparing \rf{lsn1}--\rf{lsn2} with \rf{pasta-Vir-1} one finds:
\br
\cL^{(N)}_{-1} = \pa^{\ast}_0 + Nt_1 \qquad ; \qquad 
\cL^{(N)}_0 = \pa^{\ast}_1 + N^2   \nonu  \\
\cL^{(N)}_{k-1} = \pa^{\ast}_k + (2N + k/2) \pa /\pa t_{k-1} \quad ,\;\;
k \geq 2
\lab{pasta-Vir-MM}
\er
\sect{Binary \DB Orbits on \cKPrm Hierarchies. Toda Square-Lattice Model}
\label{section:binary}
Let us recall the form of the \DB and adjoint-\DB transformations
which preserve the constrained form of \cKPrm hierarchy Lax operator \rf{f-5}
\ct{avoda,noak}, {\sl i.e.}, we shall discuss {\em auto}-\DB 
transformations for \cKPrm hierarchies (for general discussion of DB
transformations of generic KP hierarchies without the requirement of
preserving specific classes of constrained KP hierarchies, see
refs.\ct{Chau-Oevel,oevela}) :
\br
L \to {\wti L} = {\wti L}_{+} + \sum_{i=1}^m {\wti \P}_i D^{-1} {\wti \Psi}_i
= T_a L T_a^{-1} \quad ,\quad T_a = \P_a D \P_a^{-1}
\lab{DB-Lax} \\
L \to {\bar L} = {\bar L}_{+} + \sum_{i=1}^m {\bar \P}_i D^{-1} {\bar \Psi}_i
= {\bar T}^{\ast \,{-1} }_{b} L {\bar T}^{\ast}_b \quad ,\quad
{\bar T}_b = \Psi_b D \Psi_b^{-1}
\lab{adjDB-Lax}
\er
Here $\P_a ,\Psi_b$ are (adjoint) eigenfunctions entering $L_{-}$ \rf{f-5}
with fixed indices $a,b$ which henceforth will be assumed $a \neq b$. 
Accordingly, for BA functions,
the $\t$-function, eigenfunctions and their respective spectral
``densities'', the DB transformations \rf{DB-Lax} imply:
\br
{\wti \P}_a &=& T_a L (\P_a ) \quad ,\quad {\wti \Psi}_a = \frac{1}{\P_a} 
\nonu\\
{\wti \P}_i &=&  T_a (\P_i )\quad ,\quad
{\wti \Psi}_i = {T_a^{\ast}}^{-1} (\Psi_i )=  - \frac{1}{\P_a} S(\P_a ,\Psi_i )
\quad, \;\; i \neq a
\lab{DB-eigen} \\
{\wti \psi}_{BA}(t,\l ) &=&  {1\o \l} T_a (\BA ) \qquad ; \quad
{\wti \t}(t) = \P_a (t) \t (t)
\lab{DB-BA-tau} \\
{\wti \psi}^{\ast}_{BA}(t,\l ) &=&  \l {T_a^{\ast}}^{-1} (\BAc ) =
 - \l \frac{1}{\P_a (t)} S(\P_a (t) ,\BAc )
\lab{DB-BAc} \\
{\wti \vp}_a (\l ) &=&  \l^{1+r} \vp_a (\l ) \quad ;\quad
{\wti \vp}_i (\l ) = \l \vp_i (\l ) \quad ,\quad
{\wti \vp}^{\ast}_i (\l ) = \l^{-1} \vp^{\ast}_i (\l ) \quad ,\;\; i \neq a
\lab{DB-dens} 
\er
For the adjoint DB transformations \rf{adjDB-Lax} we have:
\br
{\bar \P}_b &=&  - \frac{1}{\Psi_b} \quad ,\quad
{\bar \Psi}_b = {\bar T}_b L^{\ast} (\Psi_b ) \nonu \\
{\bar \P}_i &=&  {\bar T}^{\ast \,{-1}}_{b} (\P_i )=
- \frac{1}{\Psi_b} S(\P_i ,\Psi_b ) \quad ,\quad
{\bar \Psi}_i = {\bar T}_b (\Psi_i )\quad ,\;\; i \neq b
\lab{adjDB-eigen} \\
{\bar \psi}_{BA}(t,\l ) &=&  -\l {\bar T}^{\ast \, {-1}}_{b} (\BA )
= \l \frac{1}{\Psi_b (t)} S (\BA ,\Psi_b (t) )
\lab{adjDB-BA} \\
{\bar \psi}^{\ast}_{BA}(t,\l ) &=&  - {1 \o \l} {\bar T}_b (\BAc )\qquad ;\quad
{\bar \t}(t) = \Psi_b (t) \t (t)
\lab{adjDB-BAc-tau} \\
{\bar \vp}^{\ast}_b (\l ) &=&  - \l^{1+r} \vp^{\ast}_b (\l ) \quad ; \quad
{\bar \vp}_i (\l ) = - {1\o \l} \vp_i (\l ) \quad ,\quad
{\bar \vp}^{\ast}_i (\l ) = - \l \vp^{\ast}_i (\l ) \quad , \;\; i \neq b
\phantom{aaa}
\lab{adjDB-dens}
\er
We shall use the double superscript $(n,k)$ to indicate the iteration of
$n$ successive \DB transformations \rf{DB-eigen} plus $k$ successive
adjoint-\DB transformations \rf{adjDB-eigen}. One can easily show that the
result does not depend on the particular order these transformations are
performed. Therefore, the set of all $(n,k)$ DB transformations, called
{\em generalized binary} DB transformations in what follows, defines a discrete
symmetry structure on the space of \cKPrm hierarchies corresponding to a
two-dimensional lattice. Let us note that the 
$(1,1)$ binary DB transformations were previously introduced in 
ref.\ct{oevelb}.

For one-step binary-DB transformed $\t$- and BA functions we get:
\br
\t^{(1,1)}(t) &=&  - S\(\P^{(0,0)}_a (t),\Psi^{(0,0)}_b (t)\) \t^{(0,0)}(t)
\lab{tau-11}  \\
\psi_{BA}^{(1,1)}(t,\l ) &=&  \llb 1 - {1\o \l}\frac{\P^{(0,0)}_a (t)
\Psi^{(0,0)}_b (t-[\l^{-1}])}{S\(\P^{(0,0)}_a (t),\Psi^{(0,0)}_b (t)\)}\rrb
\psi_{BA}^{(0,0)}(t,\l )   \nonu  \\
&=&  \frac{S\(\P^{(0,0)}_a (t-[\l^{-1}]),
\Psi^{(0,0)}_b (t-[\l^{-1}])\)}{S\(\P^{(0,0)}_a (t),\Psi^{(0,0)}_b (t)\)}
\psi_{BA}^{(0,0)}(t,\l )
\lab{BA-11}  \\
{\psi_{BA}^{\ast}}^{(1,1)}(t,\l ) &=& \llb 1 + {1\o \l}\frac{\Psi^{(0,0)}_b (t)
\P^{(0,0)}_a (t + [\l^{-1}])}{S\(\P^{(0,0)}_a (t),\Psi^{(0,0)}_b (t)\)}\rrb
{\psi_{BA}^{\ast}}^{(0,0)}(t,\l )   \nonu  \\
&=&  \frac{S\(\P^{(0,0)}_a (t+[\l^{-1}]),
\Psi^{(0,0)}_b (t+[\l^{-1}])\)}{S\(\P^{(0,0)}_a (t),\Psi^{(0,0)}_b (t)\)}
{\psi_{BA}^{\ast}}^{(0,0)}(t,\l )
\lab{BAc-11}
\er
where in the second equalities in \rf{BA-11} and \rf{BAc-11} we used again
\rf{whpd}--\rf{whpe} and \rf{S-id-1}--\rf{S-id-2}.
Combining eq.\rf{tau-11} with eq.\rf{tau-t-n} for $n=1$ we find the following
transformation formula for the squared eigenfunction potentials:
\be
\sum_{i=1}^m S \( \P_i^{(1,1)} , \Psi_i^{(1,1)} \)
- \sum_{i=1}^m S \( \P_i^{(0,0)} , \Psi_i^{(0,0)} \) =
\partder{}{t_r} \ln \(- S\(\P^{(0,0)}_a (t),\Psi^{(0,0)}_b (t)\)\)
\lab{DB-S}
\ee
Let us recall again that here and below $a \neq b$ are {\em fixed} indices of
Lax (adjoint) eigenfunctions.

Introducing short-hand notations:
\br
\chi_a^{(i)}(t) \equiv \( L^{(0,0)}\)^{i} (\P_a^{(0,0)}(t)) \quad ,\quad
{\chi_b^{\ast}}^{(i)}(t) \equiv {\( L^{(0,0)}\)^\ast}^{i} (\Psi_b^{(0,0)}(t))
\nonu  \\
S_{ab}^{(i,j)}(t) \equiv S\( \( L^{(0,0)}\)^{i} (\P_a^{(0,0)}(t)),
{\( L^{(0,0)}\)^\ast}^{j} (\Psi_b^{(0,0)}(t)) \)
\lab{chi-short}
\er
we have:
\begin{proposition}
The following determinant formulae hold for the $n$-step binary DB transformed
quantities:
\br
\frac{\t^{(n,n)}(t)}{\t^{(0,0)}(t)} =  (-1)^n \prod_{j=0}^{n-1}
S\( \P^{(j,j)}_a (t), \Psi^{(j,j)}_b (t)\)
= (-1)^n \Det{n}{S_{ab}^{(i-1,j-1)}(t)}
\lab{tau-nn}  \\
\phanta \nonu \\
\P_a^{(n,n)}(t) = 
(-1)^n \frac{\BDet{n+1}{S_{ab}^{(i-1,j-1)}(t)}{\chi_a^{(i-1)}
(t)}{S_{ab}^{(n,j-1)}(t)
}{\chi_a^{(n)}(t)}}{\Det{n}{S_{ab}^{(i-1,j-1)}(t)}}
\lab{P-nn} \\
\phanta \nonu \\
\Psi_a^{(n,n)}(t) = 
(-1)^n \frac{\BDet{n+1}{S_{ab}^{(i-1,j-1)}(t)}{S_{ab}^{(i-1,n)}
(t)}{{\chi_b^{\ast}}^{(j-1)}(t)}{{\chi_b^{\ast}}^{(n)}
(t)}}{\Det{n}{S_{ab}^{(i-1,j-1)}(t)}}
\lab{Psi-nn} \\
\phanta \nonu \\
S\(\P^{(n,n)}_a (t), \Psi^{(n,n)}_b (t)\) = 
\frac{\Det{n+1}{S_{ab}^{(i-1,j-1)}(t)}}{\Det{n}{S_{ab}^{(i-1,j-1)}(t)}}
\lab{S-nn}
\er
where $\chi_a^{(i)}, {\chi_b^{\ast}}^{(i)}$ and $S_{ab}^{(i,j)}$
are defined in \rf{chi-short}, and the matrix indices 
$i,j$ run from $1$ to $n$ or $n+1$ according to the indicated sizes of the
determinants.
\label{proposition:binary-DB}
\end{proposition}
Formulae \rf{P-nn}--\rf{S-nn} can be further generalized to:
\br
\( L^{(n,n)}\)^s \(\P_a^{(n,n)}(t)\) = 
(-1)^n \frac{\BDet{n+1}{S_{ab}^{(i-1,j-1)}(t)}{\chi_a^{(i-1)}
(t)}{S_{ab}^{(s+n,j-1)}
(t)}{\chi_a^{(s+n)}(t)}}{\Det{n}{S_{ab}^{(i-1,j-1)}(t)}}
\lab{L-P-nn} \\
\phanta \nonu \\  
\({L^{(n,n)}}^{\ast}\)^s \(\Psi_b^{(n,n)}(t)\) =     
(-1)^n \frac{\BDet{n+1}{S_{ab}^{(i-1,j-1)}(t)}{S_{ab}^{(i-1,s+n)}
(t)}{{\chi_b^{\ast}}^{(j-1)}(t)}{{\chi_b^{\ast}}^{(s+n)}(t)}
}{\Det{n}{S_{ab}^{(i-1,j-1)}(t)}}
\lab{L-Psi-nn} \\
\phanta \nonu \\
S\(\( L^{(n,n)}\)^l (\P^{(n,n)}_a ),
\({L^{(n,n)}}^{\ast}\)^s (\Psi^{(n,n)}_b ) \) = 
\frac{\BDet{n+1}{S_{ab}^{(i-1,j-1)}(t)}{S_{ab}^{(i-1,s+n)}
(t)}{S_{ab}^{(l+n,j-1)}
(t)}{S_{ab}^{(l+n,s+n)}(t)}}{\Det{n}{S_{ab}^{(i-1,j-1)}(t)}}
\lab{L-S-nn}
\er

We are now ready, with the help of \rf{L-P-nn}--\rf{L-S-nn},
to write down the generalization of eq.\rf{tau-nn} for the 
\cKP $\t$-function subject to an arbitrary $(n,k)$ binary DB transformation: 
\begin{proposition}
The general discrete binary \DB orbit on the space of \cKPrm
$\t$-functions, generated by a fixed pair of (adjoint) eigenfunctions
$\P_a ,\Psi_b\,$ and starting from arbitrary ``initial'' $\t^{(0,0)}$, consists
of the following elements $\t^{(n,k)}$ :
\br
\frac{\t^{(n,k)}}{\t^{(0,0)}} =
\( (-1)^k \Det{k}{S_{ab}^{(i-1,j-1)}}\)^{-(n-k-1)} \times   \nonu  \\
W_{n-k}\llb
\BDet{k+1}{S_{ab}^{(i-1,j-1)}}{\chi_a^{(i-1)}}{S_{ab}^{(k,j-1)}}{\chi_a^{(k)}},\ldots ,
\BDet{k+1}{S_{ab}^{(i-1,j-1)}}{\chi_a^{(i-1)}}{S_{ab}^{(n-1,j-1)}}{\chi_a^{(n-1)}}\rrb
\lab{tau-nm} \\
{\rm for} \;\;  n \geq k  \quad ,\quad i,j = 1,\ldots ,k   \nonu \\
\phanta \nonu \\
\frac{\t^{(n,k)}(t)}{\t^{(0,0)}(t)} =
\( (-1)^n \Det{n}{S_{ab}^{(i-1,j-1)}}\)^{-(k-n-1)} \times   \nonu  \\
W_{k-n}\llb
\BDet{n+1}{S_{ab}^{(i-1,j-1)}}{S_{ab}^{(i-1,n)}}{{\chi_b^{\ast}}^{(j-1)}
}{{\chi_b^{\ast}}^{(n)}},\ldots ,
\BDet{n+1}{S_{ab}^{(i-1,j-1)}}{S_{ab}^{(i-1,k-1)}}{{\chi_b^{\ast}}^{(j-1)}}{
{\chi_b^{\ast}}^{(k-1)}}\rrb
\lab{tau-mn}  \\
{\rm for} \;\;  n \leq k   \quad ,\quad i,j = 1,\ldots ,n   \nonu
\er
where $\chi_a^{(i)}, {\chi_b^{\ast}}^{(i)}$ and $S_{ab}^{(i,j)}$
are as in \rf{chi-short}, and $W_k \llb f_1 ,\ldots ,f_k \rrb =
\det {\Bigl\Vert} \pa^{\a -1} f_\b {\Bigr\Vert}_{\a ,\b =1,\ldots ,k}$ 
indicates Wronskian determinant of a set of functions 
$\lcurl f_1 ,\ldots ,f_k \rcurl$.
\label{proposition:binary-DB-1}
\end{proposition}
\mark
Note that the entries in the Wronskians in eqs.\rf{tau-nm}--\rf{tau-mn}
are determinants themselves.

\mark
In the Appendix we write down the explicit expressions for the \cKPrm
$\t$-functions on the most general discrete binary \DB orbit generated via
successive (adjoint) DB transformations \rf{DB-eigen}--\rf{adjDB-dens}
w.r.t. an arbitrary set of (adjoint) eigenfunctions.

In the simple case of a ``free'' initial system with $L^{(0,0)} = D$ which,
accordingly, is characterized by:
\be
\t^{(0,0)}(t) =1 \quad ; \quad
\partder{}{t_n} \chi_a^{(i)} = \pa_x^n \chi_a^{(i)}   \quad , \quad
\partder{}{t_n} {\chi_b^{\ast}}^{(j)} = - (- \pa_x )^n  {\chi_b^{\ast}}^{(j)}
\lab{tau001}
\ee
formula \rf{tau-nn} reproduces the Fredholm determinant expression
$\t = \det \left\Vert \d_{i\,j} + a_{i\,j} \right\Vert$ for the $\t$-function
with $a_{i\,j} \equiv \int_{-\infty}^x  \chi_a^{(i)}  {\chi_b^{\ast}}^{(j)} dy$
\ct{schmelz,poppe}. Namely, it follows that:
\be
\partder{}{t_n} \( \d_{ij} + a_{i\,j}\) =
\sum_{l=1}^n \pa_x^{l-1} \chi_a^{(i)} (- \pa_x )^{n-l} {\chi_b^{\ast}}^{(j)}
= {\rm Res} \( D^{-1} {\chi_b^{\ast}}^{(j)}  D^n \chi_a^{(i)} D^{-1} \)
\lab{schmelze}
\ee
The latter allows us to identify $ \d_{ij} +a_{i\,j} $ with
$S \( \chi_a^{(i)} , {\chi_b^{\ast}}^{(j)}\)$ 
and establishes connection
between the above special case of \rf{tau-nn} and the Fredholm determinant
expressions for the $\t$-functions of refs.\ct{schmelz,poppe}.

Now, we shall show that the $(n,k)$ binary \DB orbit of ${\sf cKP}_{r,2}$ 
hierarchy defines a two-dimensional Toda square-lattice 
system which describes 
two coupled ordinary two-dimensional Toda-lattice 
models corresponding to the horizontal $(n,0)$ and the vertical
$(0,k)$ one-dimensional sublattices of the $(n,k)$ binary DB square-lattice.
Namely, consider:
\be
\pa_x \partder{}{t_r} \ln \t^{(n,k)} = \Res L^{(n,k)}
= \P_1^{(n,k)} \Psi_1^{(n,k)} + \P_2^{(n,k)} \Psi_2^{(n,k)} =
\frac{\P_1^{(n,k)}}{\P_1^{(n-1,k)}} -
\frac{\Psi_2^{(n,k)}}{\Psi_2^{(n,k-1)}}
\lab{tau-P-Psi}
\ee
where we used \rf{DB-eigen} and \rf{adjDB-eigen}. Taking into account the
expressions for the (adjoint-)DB transformed $\t$-functions \rf{DB-BA-tau}
and \rf{adjDB-BAc-tau}, {\sl i.e.}
\be
\P_1^{(n,k)} = \frac{\t^{(n+1,k)}}{\t^{(n,k)}} \quad ,\quad
\Psi_2^{(n,k)} = \frac{\t^{(n,k+1)}}{\t^{(n,k)}}
\lab{P-Psi-tau}
\ee
eq.\rf{tau-P-Psi} can be rewritten in the form:
\be
\pa_x \partder{}{t_r} \ln \t^{(n,k)} = \frac{\t^{(n+1,k)} \t^{(n-1,k)}
- \t^{(n,k+1)} \t^{(n,k-1)}}{\(\t^{(n,k)}\)^2}
\lab{sq-Toda}
\ee
or, equivalently:
\be
\pa_x \partder{}{t_r} \t^{(n,k)} - \pa_x \t^{(n,k)} \partder{}{t_r} \t^{(n,k)}
= \t^{(n+1,k)} \t^{(n-1,k)} - \t^{(n,k+1)} \t^{(n,k-1)}
\lab{sq-Toda-0}
\ee
Similarly, eq.\rf{tau-P-Psi} can be represented as a system of coupled
equations of motion for $\P_1^{(n,k)}$ and $\Psi_2^{(n,k)}$ using again
\rf{P-Psi-tau} :
\br
\pa_x \partder{}{t_r} \ln \P_1^{(n,k)} =
\(\frac{\P_1^{(n+1,k)}}{\P_1^{(n,k)}} - \frac{\P_1^{(n,k)}}{\P_1^{(n-1,k)}}\)
- \(\frac{\Psi_2^{(n+1,k)}}{\Psi_2^{(n+1,k-1)}} -
\frac{\Psi_2^{(n,k)}}{\Psi_2^{(n,k-1)}}\)
\lab{sq-Toda-1} \\
\pa_x \partder{}{t_r} \ln \Psi_2^{(n,k)} =
- \(\frac{\Psi_2^{(n,k+1)}}{\Psi_2^{(n,k)}} -
\frac{\Psi_2^{(n,k)}}{\Psi_2^{(n,k-1)}}\) +
\(\frac{\P_1^{(n,k+1)}}{\P_1^{(n-1,k+1)}} -
\frac{\P_1^{(n,k)}}{\P_1^{(n-1,k)}}\)
\lab{sq-Toda-2}
\er
When $\Psi_2^{(n,k)}$ vanishes the remaining equations for  $\P_1^{(n,k)}$
reduce for a fixed $k$ to the equations of motion for the well-known
Toda model on one-dimensional lattice w.r.t. $n$ (and {\sl vice versa} if
$\P_1^{(n,k)} = 0$).

\sect{Discussion and Outlook. Relation to Random Matrix Models}
\label{section:random}
In this paper we provided a new version of the eigenfunction formulation of
KP hierarchy, called {\em squared eigenfunction potential (SEP) method},
where the SEP plays a r\^{o}le of a basic building block.
The principal ingredient of the SEP method is the proof of existence of
spectral representation for {\em any} KP eigenfunction as a spectral
integral over the (adjoint) BA function with spectral density explicitly
given in terms of a SEP.
It was pointed out that the spectral
representations of the (adjoint) BA functions themselves 
(being particular examples of KP eigenfunctions) can, in turn, serve as 
defining relations for the whole KP hierarchy parallel to Hirota fundamental 
bilinear identity or Fay identity.

The SEP method was subsequently employed to solve various issues in integrable
hierarchies of KP type both of conceptual, as well as more applied character.
Many, previously unrelated, recent developments
in the theory of $\t$-function of the KP hierarchy
gained from being described by the present formalism.
As one of the important illustrations of how our method works,
we have shown how the SEP, acting on the manifold of wave functions $\BA$
by generating non-isospectral symmetry algebra,
lifts to a vertex operator acting on $\t$-functions.
This reproduced in the SEP setting the results of 
\ct{moerbeke,ASvM,Dickey-addsym,cortona}.

We have also employed the SEP construction
in the context of Hamiltonian reductions of KP hierarchy providing:
\begin{itemize}
\item description of the reductions of the general KP hierarchy to
the constrained \cKPrm hierarchies entirely in terms of linear constraint
equations on the pertinent $\t$-functions; 
\item  description of constrained \cKPrm
hierarchies in the language of universal Sato Grassmannian; 
\item obtaining the explicit form of the non-isospectral Virasoro 
symmetry generators acting on the \cKPrm $\t$-functions. 
\end{itemize}
The achieved progress should result in further clarification of
the status of the \cKPrm hierarchies and their connection to the underlying
fermionic field language.
It would also be interesting to look for the signs of the
affine ${\hat sl} (r+m+1)$ symmetry encountered in construction
of the \cKPrm models by the generalized Drinfeld-Sokolov method 
associated to affine Kac-Moody algebras \ct{lastyear}.

Furthermore, as a principal application, the SEP method was
used to derive a series of new determinant solutions for the 
$\t$-functions of (constrained) KP hierarchies which generalize the 
familiar Wronskian (multi-soliton) solutions. These new solutions belong to a 
new type of {\em generalized binary} \DB orbits which, in turn, were shown 
to correspond to a novel Toda model on a {\em square} lattice. 
An important task for future study is to find a closed Lagrangian description 
of this new Toda square-lattice model.

Finally, let us briefly describe another potential physical application of the
present approach.

Using the spectral representation for (adjoint) eigenfunctions \rf{spec}
together with \rf{psi-main}--\rf{psi-mainc}, as well as the
following form of the Fay identity for $\t$-functions \ct{moerbeke} :
\be
\Det{n}{\frac{\t \bigl( t + [\l_i^{-1}] - [\m_j^{-1}]\bigr)}{(\l_i -\m_j )
\t (t)}} = (-1)^{{n(n-1)}\o 2}
\frac{\prod_{i>j}\(\l_i -\l_j\) \(\m_i -\m_j\)}{\prod_{i,j}\(\l_i -\m_j\)}
\frac{\t \Bigl( t + \sum_l [\l_l^{-1}] - \sum_l [\m_l^{-1}]\Bigr)}{\t (t)}
\lab{Fay-det}
\ee
we obtain an equivalent ``spectral'' representation for $\t^{(n,n)}(t)$ ~
\rf{tau-nn} :
\br
\t^{(n,n)}(t) = \frac{(-1)^{{n(n-1)}\o 2}}{(n!)^2}
\int\prod_{j=0}^{n-1}d\l_j\, d\m_j\,\prod_{i>j}\(\l_i -\l_j\)\(\l_i^r-\l_j^r\)
\prod_{j=0}^{n-1}\Bigl( {\vp^{\ast}_b}^{(0,0)}(\l_j ) e^{-\xi (t,\l_j )}\Bigr)
\times   \nonu \\
\frac{1}{{\prod_{i,j}(\l_i -\m_j )}}
\prod_{i>j}\(\m_i -\m_j\)\(\m_i^r -\m_j^r\)
\prod_{j=0}^{n-1}\Bigl( {\vp_a}^{(0,0)}(\m_j ) e^{\xi (t,\m_j )}\Bigr)\,
\t^{(0,0)} \Bigl( t + \sum_l [\l_l^{-1}] - \sum_l [\m_l^{-1}]\Bigr) \nonu \\
\phanta
\lab{tau-nn-spec}
\er

Following \ct{Yshai-Kamoto}, we can interpret the $\t$-function 
\rf{tau-nn-spec} as a partition function of certain random multi-matrix 
ensemble with the following joint distribution function of eigenvalues:
\br
Z_n \lb \{ t\} \rb \equiv const ~\t^{(n,n)}(t) =
\int\prod_{j=0}^{n-1}d\l_j\, d\m_j\, \exp\lcurl - H (t; \{\l\}, \{\m\})\rcurl
\lab{joint-distr} \\
H (t; \{\l\}, \{\m\}) \equiv \sum_j \({\bar H}_1 (\l_j ) + H_1 (\m_j )\)
+ \sum_{i>j} \( H_2 (\l_i ,\l_j ) + H_2 (\m_i ,\m_j )\)  \nonu \\
+ \sum_{i,j} {\wti H}_2 (\l_i ,\m_j ) + H_n (\{\l\}, \{\m\})
\lab{joint-H}
\er
where the one-, two- and many-body Hamiltonians read, respectively:
\br
H_1 (\l ) = - \ln \vp^{(0,0)}(\l ) - \xi (t,\l ) \quad , \quad
{\bar H}_1 (\l ) = - \ln {\vp^{\ast}}^{(0,0)}(\l ) + \xi (t,\l )
\lab{one-body} \\
H_2 (\l_i ,\l_j ) = - \ln\(\l_i -\l_j\)^2 -
\ln\Bigl( \sum_{s=0}^{r-1} \l_i^s \l_j^{r-1-s}\Bigr) \quad ,\quad
{\wti H}_2 (\l ,\m ) = \ln (\l -\m )
\lab{two-body} \\
H_n (\{\l\}, \{\m\}) =
- \ln \t^{(0,0)} \Bigl( t + \sum_l [\l_l^{-1}] - \sum_l [\m_l^{-1}]\Bigr)
\lab{many-body}
\er
The physical implications of the above new type of joint distribution function
\rf{joint-distr}--\rf{many-body} deserves further study especially regarding
critical behavior of correlations. The emerging new interesting features of
\rf{joint-distr}--\rf{many-body}, absent in the joint distribution function 
derived from the conventional two-matrix model \ct{Yshai-Kamoto}, are as 
follows:
\par\indent
(a) the second attractive term in the two-body potential $H_2$ \rf{two-body}
(both for $\l$- and $\m$- ``particles'') dominating at very long distances
over the customary repulsive first term;
\par\indent
(b) an additional two-body attractive potential ${\wti H}_2$ \rf{two-body}
between each pair of $\l$- and $\m$- ``particles''
\par\indent
(c) a genuine many-body potential $H_n$ \rf{many-body}.

One of the most important issues here is to exhibit the explicit form of the
generalized multi-matrix model behind \rf{joint-distr}--\rf{many-body}.

\appendix
\sect{Appendix: The Most General \cKPrm binary \DB Orbit.}
\label{section:appa}
Let us first introduce a few convenient compact notations for Wronskian and
related Wronskian-like determinants:
\br
W_k \equiv W_k \llb f_1 ,\ldots ,f_k \rrb =
\det {\Bigl\Vert} \pa^{\a -1} f_\b {\Bigr\Vert} \quad , \quad
W_k (f) \equiv W_{k+1} \llb f_1 ,\ldots ,f_k , f \rrb
\lab{wronski-def} \\
{\wti W}_{k+1}(f) \equiv {\wti W} \llb f_1 ,\ldots ,f_{k+1}; f \rrb \equiv
\BDet{k+1}{\pa^{\a -1}f_\b}{\pa^{\a -1}f_{k+1}}{S\( f_\b ,f\)}{S\( f_{k+1},f\)}
\lab{wronski-bar-def}
\er
where the matrix indices $\a ,\b =1,\ldots ,k$ and, as above,
$\pa_x S\( f_\b ,f \) = f_\b\, f$ . The Wronskian\-(-like) determinants 
\rf{wronski-def}--\rf{wronski-bar-def} obey the following useful identities:
\be
\pa \(\frac{W_{k-1}(f)}{W_k}\) = \frac{W_k (f) \, W_{k-1}}{W_k^2} 
\quad ; \quad
\pa \(\frac{{\wti W}_{k+1}(f)}{W_k}\) = 
- \frac{{\wti W}_k (f)\, W_{k+1}}{W_k^2} 
\lab{Jacobi-bar-id}
\ee
where the first one is known as Jacobi expansion theorem (see, {\sl e.g.}
\ct{Wronski}), whereas the second identity in \rf{Jacobi-bar-id} can be easily
verified via induction. Eqs.\rf{Jacobi-bar-id} imply in turn the identities:
\be
T_k \cdots T_1 (f) = \frac{W_k (f)}{W_k} \quad ,\quad
{T^{\ast}_k}^{-1} \cdots {T^{\ast}_1}^{-1} (f) = - \frac{{\wti W}_k (f)}{W_k}
\lab{iw-bar}
\ee
with:
\be
T_j = {W_{j} \over W_{j-1}} D {W_{j-1} \over W_{j}}  \quad , \quad
{T^{\ast}_j}^{-1} = - {W_{j-1}\over W_{j}} D^{-1} {W_{j} \over W_{j-1}} 
\ee

Now we can use eqs.\rf{iw-bar} to derive explicit expressions for the
(adjoint) eigenfunctions and $\t$-functions of \cKPrm hierarchies generated
via successive (adjoint) DB transformations \rf{DB-eigen}--\rf{adjDB-dens}
w.r.t. an arbitrary set of (adjoint) eigenfunctions of the ``initial'' \cKPrm 
Lax operator $L = L_{r,m}$ \rf{f-5}.
We shall denote the latter arbitrary successive (adjoint) DB transformations
by the following double-vector superscript:
\be
\(\vec{n},\vec{k}\) \equiv \( (n_1 ,\ldots ,n_m ), (k_1 ,\ldots k_m )\)
\lab{vec-supscript}
\ee
indicating $n_1$ successive DB transformations w.r.t. $\P_1$ etc., until
$n_m$ DB transformations w.r.t. $\P_m$ and,
similarly, $k_1$ successive adjoint-DB transformations w.r.t. $\Psi_1$ etc.,
until $k_m$ adjoint-DB transformations w.r.t. $\Psi_m$. Specifically, we have:
\br
\P_a^{\(\vec{n},\vec{0}\)} = (-1)^{\sum_{a+1}^m n_j} \,
\frac{W\llb \chi^{(0)}_1, \ldots ,\chi^{(n_1-1)}_1, \ldots , 
\chi^{(0)}_a, \ldots ,\chi^{(n_a-1)}_a, \chi^{(n_a)}_a, 
\ldots , \chi^{(0)}_m, \ldots ,
\chi^{(n_m-1)}_m \rrb}{W\llb \chi^{(0)}_1, \ldots ,\chi^{(n_1-1)}_1, \ldots , 
\chi^{(0)}_a, \ldots ,\chi^{(n_a-1)}_a, 
\ldots , \chi^{(0)}_m, \ldots ,\chi^{(n_m-1)}_m \rrb}  \nonu \\ 
\phantom{aaa} \lab{pchi-a-1}  \\
\frac{\t^{\(\vec{n},\vec{0}\)}}{\t^{\(\vec{0},\vec{0}\)}} =
W\llb \chi^{(0)}_1, \ldots ,\chi^{(n_1-1)}_1, \ldots , 
\chi^{(0)}_m, \ldots ,\chi^{(n_m-1)}_m \rrb   \phantom{aaaaaaaaaaaaaa}
\lab{tauok-1}  \\
\Psi_a^{\(\vec{n},\vec{0}\)} = \left\{ \begin{array}{ll}
(-1)^{\sum_{a+1}^m n_j}
\frac{W\llb \chi^{(0)}_1, \ldots ,\chi^{(n_1-1)}_1, \ldots , 
\chi^{(0)}_a, \ldots ,\chi^{(n_a-2)}_a , 
\ldots , \chi^{(0)}_m, \ldots ,
\chi^{(n_m-1)}_m \rrb}{W\llb \chi^{(0)}_1, \ldots ,\chi^{(n_1-1)}_1, \ldots , 
\chi^{(0)}_a, \ldots ,\chi^{(n_a-1)}_a, 
\ldots , \chi^{(0)}_m, \ldots ,\chi^{(n_m-1)}_m \rrb}
\quad &, \quad {\rm for} \;\; n_a \geq 1  \\
\phantom{aaa}   \\
- \frac{{\wti W} \llb \chi^{(0)}_1, \ldots ,\chi^{(n_1-1)}_1, \ldots , 
\chi^{(0)}_m, \ldots ,\chi^{(n_m-1)}_m ; \Psi_a^{\(\vec{0},\vec{0}\)}  
\rrb}{W\llb \chi^{(0)}_1, \ldots ,\chi^{(n_1-1)}_1, \ldots , 
\chi^{(0)}_m, \ldots ,\chi^{(n_m -1)}_m \rrb}
\quad &, \quad {\rm for} \;\; n_a = 0    
\end{array} \right.
\lab{psichi-a-1}
\er
where the functions $\chi^{(s)}_a$ are the same as in \rf{chi-short} with the
superscripts $(0,0)$ replaced with the corresponding double-vector ones
$\(\vec{0},\vec{0}\)$. Eqs.\rf{pchi-a-1}--\rf{tauok-1} already appeared in
\ct{addsym} (see also refs.\ct{pirin-wilox}), whereas eq.\rf{psichi-a-1} is
derived via iterative application of the second identity in \rf{iw-bar} and
taking into account \rf{pchi-a-1}.

Now, performing arbitrary successive adjoint-DB transformations on
$\t^{\(\vec{n},\vec{0}\)}$ \rf{tauok-1} according to the second
eq.\rf{adjDB-BAc-tau} upon using the first identity in \rf{iw-bar} and
inserting there the explicit expressions \rf{psichi-a-1}, we arrive at the
following:
\begin{proposition}
The most general discrete binary \DB orbit on the space of \cKPrm
$\t$-functions is built-up of the following elements:
\br
\frac{\t^{\(\vec{n},\vec{k}\)}}{\t^{\(\vec{0},\vec{0}\)}} =
\( - W \llb \chi^{(0)}_1, \ldots ,\chi^{(n_1-1)}_1, \ldots , \chi^{(0)}_a, 
\ldots ,\chi^{(n_a-1)}_a \rrb\)^{-\sum_{a+1}^m k_j} \times \nonu \\
W\llb \D^{\vec{n}}_{(0,a+1)},\ldots ,\D^{\vec{n}}_{(k_{a+1}-1,a+1)},\ldots ,
\D^{\vec{n}}_{(0,m)},\ldots ,\D^{\vec{n}}_{(k_m -1,m)} \rrb
\lab{tau-nk-gen} \\
\phantom{aaa} \nonu \\
\D^{\vec{n}}_{(l,s)} \equiv 
{\wti W} \llb \chi^{(0)}_1, \ldots ,\chi^{(n_1-1)}_1, \ldots , \chi^{(0)}_a, 
\ldots ,\chi^{(n_a-1)}_a; {\chi_s^{\ast}}^{(l)} \rrb
\lab{Delta-def}
\er
where:
\be
\(\vec{n},\vec{k}\) = 
\( (n_1 ,\ldots ,n_a ,0,\ldots ,0), (0,\ldots ,0,k_{a+1},\ldots ,k_m )\) 
\quad ;\quad  a=0,1,\ldots ,m   
\lab{DB-ranges}
\ee
and, furthermore, notations \rf{chi-short} and \rf{wronski-bar-def} are 
employed.
\label{proposition:binary-DB-2}
\end{proposition}
\mark
The reason for the zero entries in the labels \rf{DB-ranges} of the most 
general binary DB transformations, preserving the spaces of \cKPrm hierarchies
\rf{f-5}, lies in the fact that any pair of two successive (adjoint-)DB
transformations w.r.t. $\P_a ,\Psi_a$, {\sl i.e.} both with the {\em same} 
index, is equivalent to an identity transformation as one can easily conclude 
by combining the second equation in \rf{DB-eigen} with the second equations in
\rf{DB-BA-tau} and \rf{adjDB-BAc-tau}.

\end{document}